%% file: paper.tex
\documentclass[12pt,letterpaper]{article}
\usepackage[letterpaper, total={7in, 9in}]{geometry}
\usepackage[font=footnotesize,labelfont=bf]{caption}
\usepackage{graphicx}%
\usepackage{multirow}%
\usepackage{authblk}
\usepackage{orcidlink}
\usepackage{amsmath,amssymb,amsfonts}%
\usepackage{amsthm}
\usepackage{url}
\usepackage{mathrsfs}%
\usepackage{textalpha}
\usepackage[title]{appendix}%
\usepackage[version=4]{mhchem}
\usepackage{xr}
\usepackage{xcolor}%
\usepackage{textcomp}%
\usepackage{manyfoot}%
\usepackage{booktabs}%
\usepackage{algorithm}%
\usepackage{algorithmicx}%
\usepackage{algpseudocode}%
\usepackage{listings}%
\usepackage{setspace} %
\usepackage{soul}   
\usepackage{hyperref}
\hypersetup{
  colorlinks   = true, 
  urlcolor     = blue, 
  linkcolor    = blue, 
  citecolor    = blue, 
}
\usepackage[super,comma,sort&compress]  
   {natbib}
\usepackage[right]{lineno} 

\usepackage{pdfpages}
\usepackage{pgffor} 

\makeatletter
\renewcommand{\maketitle}{\bgroup\setlength{\parindent}{0pt}
\begin{flushleft}
  {\LARGE\textbf{\@title}}
  \vspace{1em}
  
  \@author
\end{flushleft}\egroup}
\makeatother

\makeatletter
\newcommand*{\addFileDependency}[1]{
\typeout{(#1)}
\@addtofilelist{#1}
\IfFileExists{#1}{}{\typeout{No file #1.}}
}
\makeatother

\newcommand*{\myexternaldocument}[1]{%
\externaldocument{#1}%
\addFileDependency{#1.tex}%
\addFileDependency{#1.aux}%
}

\myexternaldocument{SI}

  {\vspace*{\fill}\clearpage\restoregeometry}

\makeatletter
\AtBeginDocument{\let\LS@rot\@undefined}
\makeatother

\def\supplementfilename{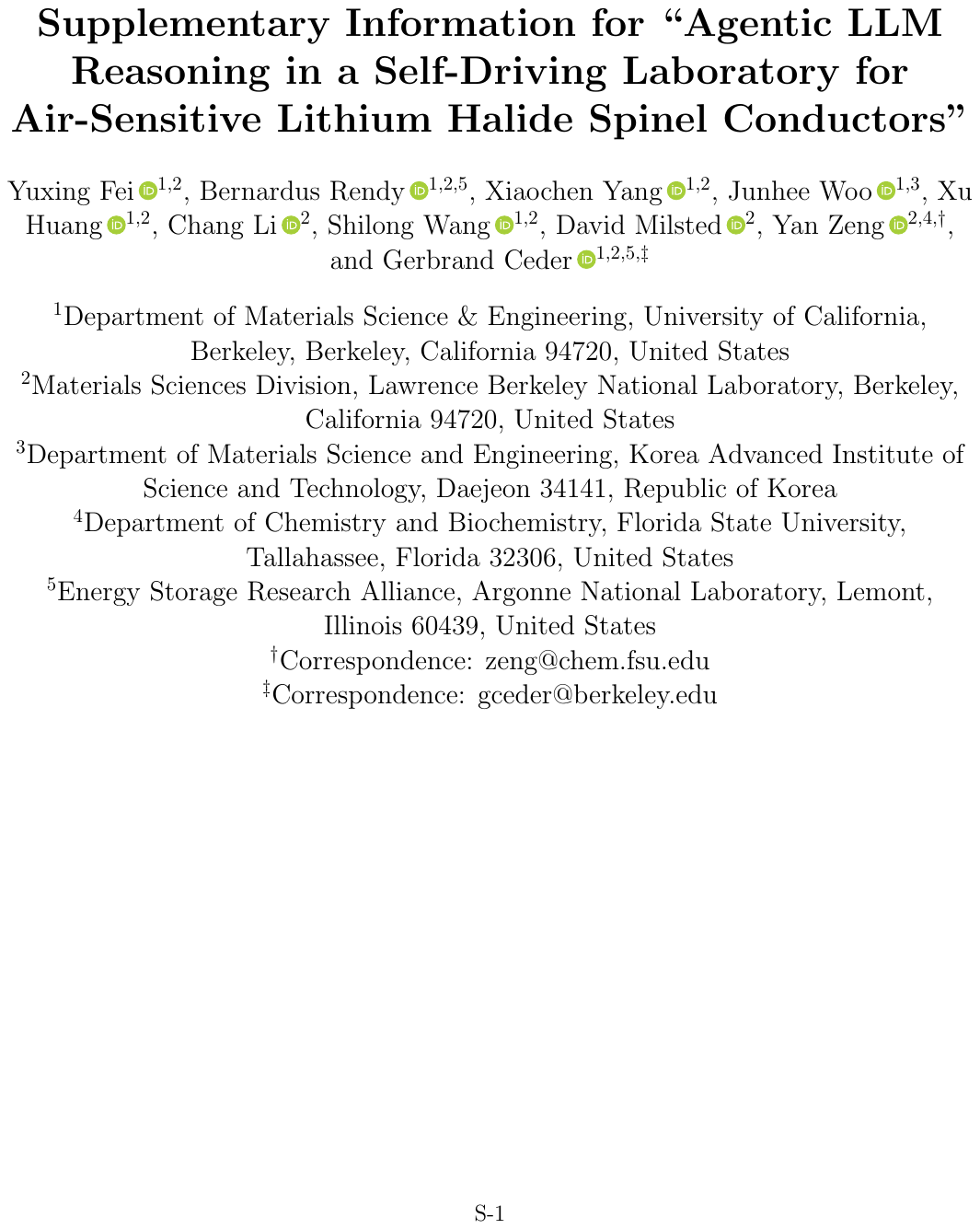}

\pdfximage{\supplementfilename}
\def\numbersupplementpages{\the\pdflastximagepages}

\newif\ifarXiv
\arXivtrue

\begin{document}

\title{Agentic LLM Reasoning in a Self-Driving Laboratory for Air-Sensitive Lithium Halide Spinel Conductors}
\author[1,2]{Yuxing Fei\,\orcidlink{0000-0002-1225-2083}}
\author[1,2,5]{Bernardus Rendy\,\orcidlink{0000-0001-8309-6279}}
\author[1,2]{Xiaochen Yang\,\orcidlink{0000-0002-8359-5630}}
\author[1,3]{Junhee Woo\,\orcidlink{0009-0002-1869-2861}}
\author[1,2]{Xu Huang\,\orcidlink{0009-0002-2260-5150}}
\author[2]{Chang Li\,\orcidlink{0000-0001-5420-3856}}
\author[1,2]{Shilong Wang\,\orcidlink{0009-0004-8504-5802}}
\author[2]{David Milsted\,\orcidlink{0000-0003-0415-910X}}
\author[2,4,$\dagger$]{Yan Zeng\,\orcidlink{0000-0002-5831-1210}}
\author[1,2,5,$\ddagger$]{Gerbrand Ceder\,\orcidlink{0000-0001-9275-3605}}

\affil[1]{Department of Materials Science \& Engineering, University of California, Berkeley, Berkeley, California 94720, United States}

\affil[2]{Materials Sciences Division, Lawrence Berkeley National Laboratory, Berkeley, California 94720, United States}

\affil[3]{Department of Materials Science and Engineering, Korea Advanced Institute of Science and Technology, Daejeon 34141, Republic of Korea}

\affil[4]{Department of Chemistry and Biochemistry, Florida State University, Tallahassee, Florida 32306, United States}

\affil[5]{Energy Storage Research Alliance, Argonne National Laboratory, Lemont, Illinois 60439, United States}
\affil[$\dagger$]{Correspondence: zeng@chem.fsu.edu}
\affil[$\ddagger$]{Correspondence: gceder@berkeley.edu}
\setstretch{1.1} 

\maketitle
\section*{Summary}
Self-driving laboratories promise to accelerate materials discovery. Yet current automated solid-state synthesis platforms are limited to ambient conditions, thereby precluding their use for air-sensitive materials. Here, we present A-Lab for Glovebox Powder Solid-state Synthesis (A-Lab GPSS), a robotic platform capable of synthesizing and characterizing air-sensitive inorganic materials under strict air-free conditions. By integrating an agentic AI framework into the A-Lab GPSS platform, we structure autonomous experimental design through abductive and inductive reasoning. We deploy this platform to explore the vast compositional space of lithium halide spinel solid-state ionic conductors. Across a synthesis campaign comprising 352 samples with diverse compositions, the system explores a broad chemical space, experimentally realizing 72\% of the 171 possible pairwise combinations among the 19 metals considered in this study. Over the course of the campaign, the fraction of compositions exhibiting both good ionic conductivity (\textgreater 0.05 mS/cm) and high halide spinel phase purity increases from 1.33\% in the first 75 agent-proposed samples to 5.33\% in the final 75. Furthermore, by inspecting the AI's reasoning processes, we reveal distinct yet complementary discovery strategies: abductive reasoning interrogates abnormal observations within already explored regions, whereas inductive reasoning expands the search into broader, previously unvisited chemical space.
This work establishes a scalable platform for the autonomous discovery of complex, air-sensitive solid-state materials.
\section*{Keywords}
self-driving laboratory; solid-state synthesis; lithium halide spinel; ionic conductors; large language model agents; scientific reasoning

\section{Introduction}
Data-driven, AI-guided computational discovery has rapidly emerged as a new paradigm in materials science. \cite{himanen2019data, otyepka2025advancing, jain2013commentary, horton2025accelerated, cheng2026artificial, saal2013materials} Yet, as materials science is grounded in physical reality, computational predictions must be validated experimentally through synthesis and characterization in the laboratory. This creates a fundamental bottleneck, as AI-driven design can generate candidate materials far more rapidly than traditional laboratory workflows can evaluate them. \cite{merchant2023scaling}

Self-driving laboratories (SDLs) offer a promising solution to this throughput mismatch.
By combining robotic automated sample and hardware handling with AI-driven data analysis and decision-making, self-driving laboratories create closed-loop experiment pipelines that substantially increase throughput, enhance sample-to-sample reproducibility, and generate high-fidelity, provenance-rich datasets suited for downstream AI-driven analysis. These platforms have been successfully deployed across a wide range of chemical and material systems, including solid-state crystalline powders \cite{szymanski2023autonomous, chen2024navigating}, nanomaterials/quantum dots \cite{bateni2024smart, zaki2025self}, polymers \cite{wang2025autonomous}, electrolyte formulation \cite{yik2025accelerating, zhang2025salsa}, catalysts \cite{burger2020mobile, tinajero2025reac, zhang2025multimodal}, thin films \cite{macleod2020self, macleod2022self, cakan2024pascal, ament2021autonomous}, organic molecules \cite{steiner2019organic, ruan2024automatic}, alloys \cite{ghafarollahi2025automating}, carbon capture materials \cite{giro2023ai}, and more \cite{tom2024self}.

Despite rapid progress in a broad range of areas, self-driving laboratories for solid-state synthesis remain comparatively underexplored\cite{lo2024review}, even though solid-state reactions are commonly used for the synthesis of inorganic materials. This gap stems from the practical challenges of high-temperature processing and reliable powder handling, which complicate the design and operation of autonomous systems. In addition, existing solid-state synthesis autonomous laboratories operate under ambient conditions, limiting their applicability to air-stable compounds and excluding technologically vital materials, such as halides, chalcogenides, and pnictogenides. Addressing these challenges presents an important opportunity to extend autonomous discovery of inorganic materials to a broader chemical space and to enable systematic exploration of air-sensitive inorganic systems via a solid-state route.

Herein, we present A-Lab for Glovebox Powder Solid-state Synthesis (A-Lab GPSS), an integrated, autonomous platform for the solid-state synthesis of air-sensitive materials, with semi-automated characterization workflows for powder X-ray diffraction (XRD) and electrochemical measurements. The experiment system is connected to an AI agent system for closed-loop experimental design and hypothesis generation, which consists of AI agents acting in two complementary modes of scientific reasoning: (1) \textit{abductive reasoning}, which generates hypotheses to explain unexpected observations, and (2) \textit{inductive reasoning}, which distills patterns across accumulated data.
We showcase this self-driving system in a synthesis campaign targeting lithium halide spinel solid-state ionic conductors. With the help of the A-Lab GPSS and state-of-the-art LLM agents, 352 samples were synthesized and characterized by XRD and ionic conductivity measurements. Our analysis shows that this self-driving system can cover a wide range of chemical space, encompassing 19 metals (21 cation species, including multiple valence states for some metals), spanning 72\% of the 171 possible pairwise metal combinations. Over the course of the campaign, the success rate, defined as the fraction of samples exhibiting both good ionic conductivity ($>0.05$ mS/cm) and high spinel purity, increased by approximately fourfold, from 1.33\% in the first 75 agent-proposed samples to 5.33\% in the final 75. Further analysis reveals that the agents adopt distinct yet complementary strategies, forming a synergistic discovery framework. Abductive reasoning focuses on anomalous observations within already explored regions, enabling finer re-exploration of chemical space, whereas inductive reasoning drives broader exploration of previously unvisited chemical space. 

\section{Results}
\subsection{Automated solid-state powder synthesis system for air-sensitive materials}
\begin{figure}
    \centering
    \includegraphics[width=0.9\linewidth]{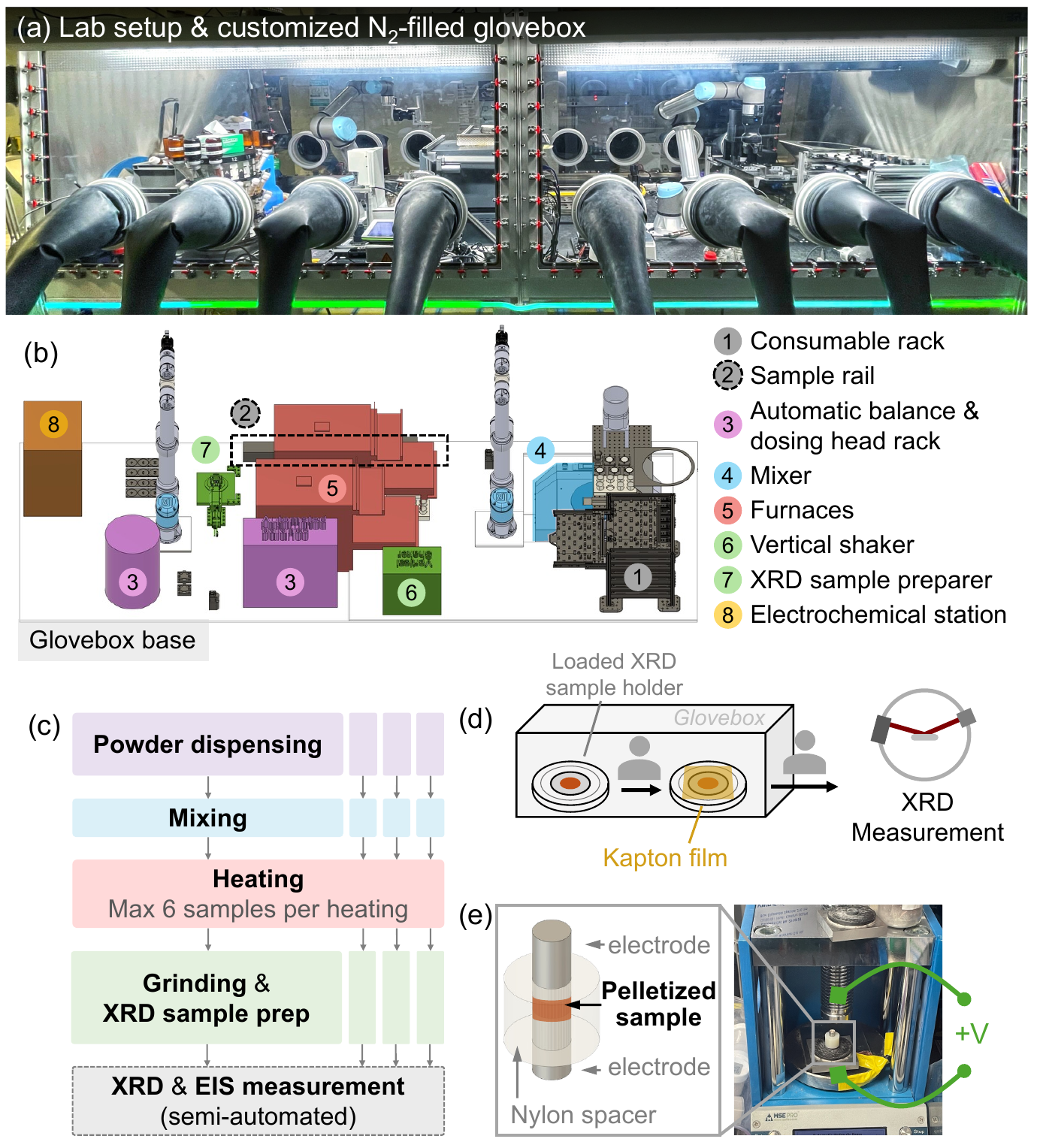}
    \caption[Overview of the A-Lab GPSS platform for air-sensitive solid-state synthesis, characterization, and impedance measurement.]{\textbf{Overview of the A-Lab GPSS platform for air-sensitive solid-state synthesis, characterization, and impedance measurement.}
    (a) Photograph of the customized \ce{N2}-filled glovebox system housing the A-Lab GPSS setup. 
    (b) Three-dimensional schematic of the GPSS glovebox layout, with individual modules color-coded and labeled on the right. 
    (c) Synthesis and characterization workflow, comprising automated powder dispensing, mixing, high-temperature heating, grinding, and XRD sample preparation, followed by semi-automated powder XRD and electrochemical impedance spectroscopy (EIS) measurements. Multiple samples are connected to the heating tasks, as one heating cycle can handle multiple samples (up to 6).
    (d) Manual transfer step for sealing prepared XRD samples with Kapton film prior to XRD measurement outside the glovebox. 
    (e) Customized electrochemical station for EIS measurements. The hydraulic press (right) is electrically connected to an external electrochemical workstation. During pressing, the pressing rod and the plate form a closed electrical circuit. The inset (left) shows the custom test cell, in which the sample is pelletized between two stainless-steel rod electrodes.}
    \label{fig:fig1}
\end{figure}

The A-Lab for Glovebox Powder Solid-state Synthesis (A-Lab GPSS) is an integrated robotic platform operated under a nitrogen-filled, air-free environment for the synthesis, characterization, and testing of materials. Figure \ref{fig:fig1}a shows an overview of the A-Lab GPSS setup enclosed in a customized double-glovebox. Inside the glovebox, there are five integrated workstations to realize an automated solid-state synthesis pipeline, including precursor powder dispensing, mixing, heating, post-heating grinding, and XRD sample preparation. Two robot arms mounted on the glovebox floor perform different operations and manipulate the samples, including a linear rail for sample transfer. Figure \ref{fig:fig1}b details the spatial arrangement of the robotic arms and workstations.

Beyond automated synthesis, A-Lab GPSS integrates a semi-automated characterization workflow comprising powder X-ray diffraction (XRD; Figure \ref{fig:fig1}d) and electrochemical impedance spectroscopy (EIS; Figure \ref{fig:fig1}e). Samples prepared for XRD measurement are manually sealed with Kapton films (shown in Figure \ref{fig:fig1}d), which are then transferred to the X-ray diffractometer outside the glovebox. For EIS measurements, samples are handled inside the glovebox the entire time. Specifically, powder samples are loaded into a custom-built test cell (Figure \ref{fig:fig1}e, left), where the sample is sandwiched between two stainless-steel rod electrodes and constrained by a nylon spacer. The assembled cell is placed in an electric hydraulic press, which simultaneously pelletizes the powder and serves as the measurement fixture. Impedance spectra are collected by electrically connecting the potentiostat leads to the upper pressing rod and the lower base plate. The acquired spectra are analyzed by fitting to an equivalent-circuit model to extract the ionic conductivity. Full details of the circuit definition and fitting procedure are provided in the Methods section. It is worth noting that while a fully automated, end-to-end synthesis and characterization pipeline is technically feasible, practical considerations, including cost, robustness, and experimental flexibility, lead to the implementation of a combination of automated synthesis and semi-automated post-synthetic handling in A-Lab GPSS. \cite{mccalla2023semiautomated}

A key design challenge in integrating the full solid-state synthesis pipeline within a glovebox is maximizing space utilization without compromising operability. As the glovebox is a constrained space, equipment must be packed densely while maintaining sufficient clearance for robotic motion and occasional manual intervention. We address this primarily through a system-wide vertical ``stacking'' strategy. For example, the mixer (labeled \textcircled{4} in Figure \ref{fig:fig1}b) is seated in a custom subfloor well, as the robot arm only needs to load/unload samples from the top of the machine. Above the mixer, a secondary platform is built to provide an intermediate staging location for samples being processed. Another example is the consumable rack labeled as \textcircled{1} in Figure \ref{fig:fig1}b. All the consumables (crucible, plastic vial, and caps) are stored on a seven-tier rack. If needed, each rack can be pulled out by the robot as a drawer to retrieve consumables. Together, these design elements increase space efficiency by exploiting the glovebox's vertical volume. Because the glovebox cannot be opened during normal operation, the majority of equipment is positioned along the sidewalls for access through glove ports, whereas lower-maintenance, off-the-shelf components, such as the robot arms and furnaces (labeled \textcircled{5} in Figure \ref{fig:fig1}b), are placed along the glovebox centerline, where manual access is more limited but the space can be used more efficiently.

The A-Lab GPSS platform is generally suitable for synthesizing and processing materials via the solid-state route which are sensitive to air, \ce{O2}, \ce{CO2}, \ce{H2O}, etc. In the present study, we apply the A-Lab GPSS platform to the synthesis and exploration of halide materials, many of which are known to be extremely sensitive to trace amounts of \ce{H2O} \cite{wang2022air, ren2025humidity}. Figure \ref{fig:fig1}c presents a typical experimental flowchart. The average processing time per sample is approximately 19 minutes for powder dispensing, 20 minutes for mixing, and 37 minutes for grinding and XRD sample preparation after heating. The heating step is the primary rate-limiting step, requiring an average of 21 hours per heating cycle. To mitigate this bottleneck, two furnaces are operated in parallel in A-Lab GPSS, and each furnace can process up to six samples concurrently when these samples share an identical heating profile. Considering that many halides exhibit low melting points and non-negligible vapor pressures at elevated temperatures, each crucible is covered with a metal lid to reduce volatilization and minimize cross-contamination during heating. (Supplementary Figure \ref{si-fig:crucible-cap}, Supplementary Note \ref{si-note:consumbles}) A detailed description of the automated experimental workflow is provided in the Methods section and is shown in the Supplementary Video.

\subsection{Agentic workflows for experimental design}
Large language model (LLM)-based agents have been used as general-purpose decision-makers to discover new materials with desired properties. \cite{zimmermann202532} With their ability to ingest and reason over vast corpora of scientific knowledge, these agents are increasingly deployed in self-driving laboratories, where they query previous experimental data, run simulations, and command robotic laboratories to autonomously design and execute experiments. \cite{boiko2023autonomous, zhou2026practical, panapitiya2025autolabs, gupta2025llms, cao2025automating, huang2025cascade, m2024augmenting}
Most existing systems employ LLM agents in a monolithic, end-to-end fashion: agents review prior outcomes, invoke external tools, refine designs, and propose subsequent experiments. From a scientific standpoint, this workflow conflates two fundamentally distinct modes of inference \cite{smith2022types_inferences}: abductive reasoning (generating plausible hypotheses to explain observations) and inductive reasoning (distilling regularities from accumulated data). Consequently, the internal logic flow of these agents remains untraceable. \cite{wang2025agentarmor} It is often unclear whether a successful design stems from hypothesis-driven abductive reasoning, pattern-based induction, or merely superficial correlations inherited from the training data. This ambiguity hinders a rigorous assessment of the agents' true scientific capabilities and their failure modes, especially when navigating uncharted material spaces.

\begin{figure}
    \centering
    \includegraphics[width=0.9\linewidth]{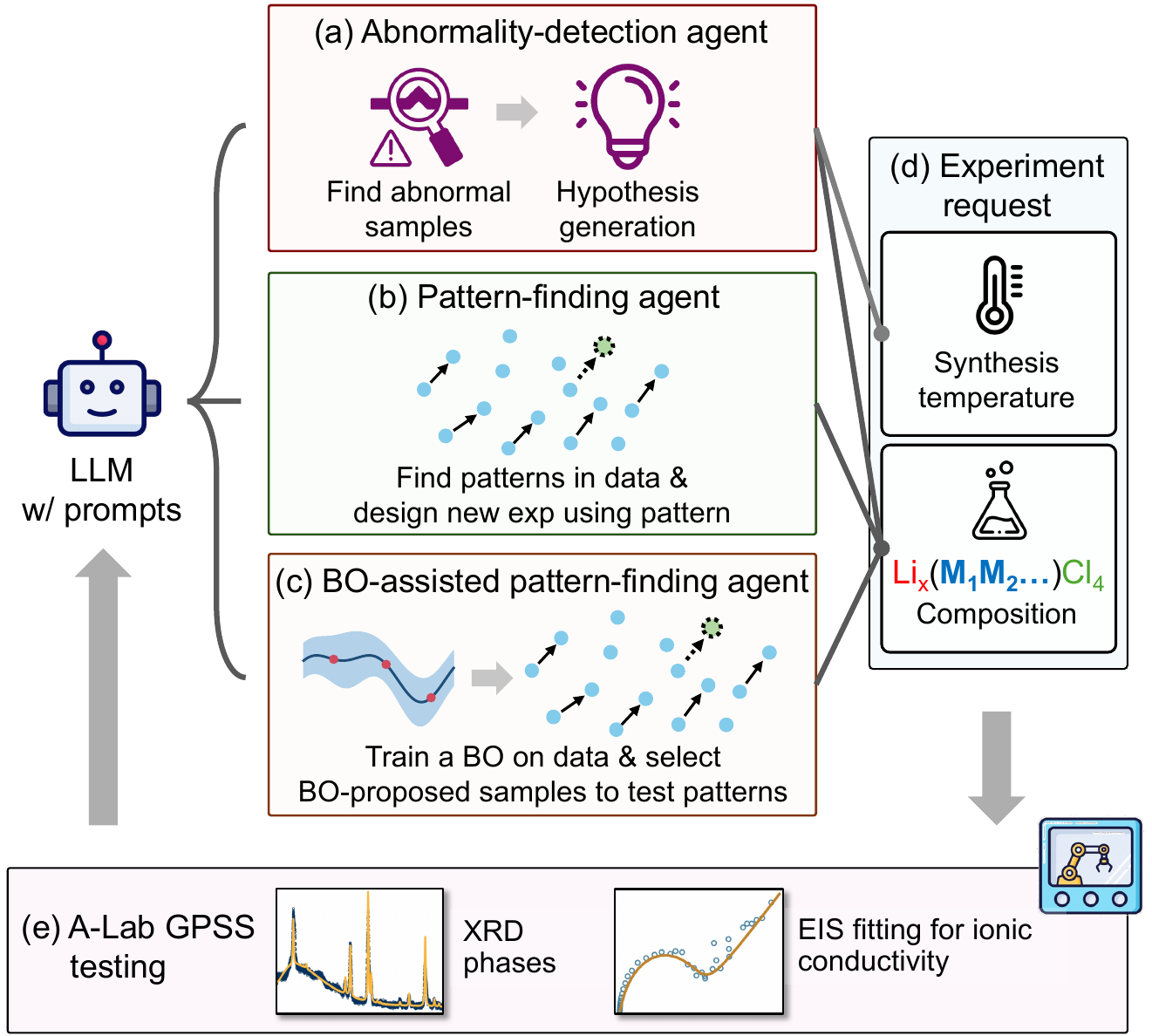}
    \caption[Agentic workflow for experimental design.]{\textbf{Agentic workflow for experimental design.}
(a-c) Experimental design agent workflows, each with different system prompts to focus on different aspects of scientific reasoning, using the same underlying LLM (OpenAI GPT-5).
(a) The abnormality-detection agent first calls the LLM to analyze existing experimental data to identify anomalous samples whose XRD-identified phases or ionic conductivity deviate significantly from other chemically similar samples. For each detected anomaly, a subsequent LLM call generates a hypothesis explaining the deviation and proposes a targeted follow-up experiment to test the hypothesis.
(b) The pattern-finding agent compares multiple samples to extract recurring trends and correlations in the dataset, and designs new experiments by extrapolating or applying these learned patterns.
(c) The Bayesian-optimization- (BO-)assisted pattern-finding agent extends the pattern-finding workflow by first training a BO model on the full dataset to propose candidate compositions. These candidates are then evaluated by the LLM, which selects compositions that are consistent with the identified experimental patterns and introduces modifications only when necessary.
(d) Agent outputs include all key parameters required for experiment execution. For the abnormality-detection agent, both the target composition and synthesis temperature may be adjusted. For the two pattern-finding agents, only the target composition is modified, while the synthesis temperature is estimated based on the melting points of the precursor materials.
(e) The proposed samples are subsequently sent to A-Lab GPSS for experimental validation. The resulting XRD phase information and ionic conductivity are analyzed and fed back to the agents, which then run the next experimental design cycle.} 
    \label{fig:fig2}
\end{figure}

Motivated by this scientific-reasoning perspective, we design three LLM agentic workflows to operate on real-world, often noisy, experimental datasets generated in A-Lab GPSS (Figure \ref{fig:fig2}). Rather than treating LLMs as end-to-end decision-makers, we structure the agents into two complementary modes of scientific reasoning: (1) \textit{abductive reasoning}, which generates hypotheses and new experiments to explain unexpected observations, and (2) \textit{inductive reasoning}, which distills patterns across accumulated data. We implement three coordinated agents, an abnormality-detection (abductive) agent and two pattern-finding (inductive) agents, including a Bayesian-optimization-assisted variant, to drive selection on experimental parameters. This structured decomposition enables some traceability of agents' decision-making process: each proposed experiment can be traced to hypothesis-driven probing of samples with abnormal outcomes or to pattern-driven design of samples with emerging trends in the experimental dataset.

\subsubsection*{Abnormality-detection agent}
The abnormality-detection agent (Figure \ref{fig:fig2}a) is designed for abductive reasoning on the experimental data, with a focus on understanding the causes of abnormal samples in the dataset. Abnormalities here are defined as either the emergence of unexpected phases in the XRD patterns or measured ionic conductivity values that deviate substantially from those of chemically similar samples, typically with comparable elemental compositions and stoichiometries. This agent is prompted by an LLM call to search for abnormalities in the experimental dataset provided as input. Once an abnormal sample is identified, another LLM call is made to generate a hypothesis for the possible cause of the abnormality. This agent is designed to reflect a common scenario in experimental materials research: an unexpected result prompts the researcher to propose an informed hypothesis about its origin and to design targeted follow-up experiments to test that hypothesis, potentially leading to new insights for material design.

\subsubsection*{Pattern-finding agent}
The pattern-finding agent (Figure \ref{fig:fig2}b) is designed to emphasize inductive reasoning over accumulated experimental evidence. Instead of focusing on individual outliers, this agent jointly analyzes multiple samples to extract trends in phase purity and ionic conductivity. It is prompted to summarize potential patterns and propose new experiments by systematically applying the observed trends to unexplored regions of composition space. In this mode, experimental design is driven by the identification of regularities that suggest generalizable design principles without the need to understand a specific failure or anomaly. This agent mirrors a common inductive workflow in traditional materials research, where empirical rules, such as favorable cation combinations or compositional motifs associated with improved ionic conductivity, are distilled from prior results and then used to guide subsequent exploration.

\subsubsection*{BO-assisted pattern-finding agent}
The Bayesian optimization (BO)-assisted pattern-finding agent (Figure \ref{fig:fig2}c) is utilized at a later stage of the experimental campaign, when a sufficiently large and diverse dataset has been accumulated to support data-driven surrogate modeling. The BO-assisted agent is used in combination with the pattern-finding agent to tackle the difficulties in effective learning and reasoning with an increasing sample size in the dataset. In this workflow, a BO model is first trained on all explored samples to propose 40 candidate compositions predicted to have high spinel phase content and ionic conductivity, while remaining unexplored (high uncertainty). The details of training the BO model are provided in the Supplementary Note \ref{si-note: bo-training}.  These BO-suggested candidates are then passed to the LLM pattern-finding agent, which evaluates them in the context of identified experimental patterns. The agent selects a subset of the BO proposals, potentially introducing minimal modifications (e.g., removing one element or slightly adjusting the stoichiometry) to maximize alignment with the patterns inferred from the experimental dataset. In this way, the BO model explicitly constrains the pattern-finding agent's search space by providing a focused set of high-value candidates, enabling the LLM to focus its reasoning on compositions that are both promising and aligned with the trend revealed in the explored data.
\\\\
The tunable experimental parameters available to each agent differ by design (Figure \ref{fig:fig2}d). The abnormality-detection agent is permitted to adjust both the target composition and the synthesis temperature. In contrast, the two pattern-finding agents are restricted to proposing compositions only; their synthesis temperatures are automatically estimated using a modified Tammann's rule approximation \cite{merkle2005tammann}, set to three-quarters of the average melting point of the precursors. This design reflects different objectives of agents. Anomalous outcomes in phase purity or ionic conductivity often arise from temperature-dependent effects, such as incomplete solid-state reactions or slow kinetics, in which case adjusting the synthesis temperature can yield informative results. By contrast, the pattern-finding agents are encouraged to focus on identifying promising composition-property relationships. Allowing temperature adjustment would substantially expand the design space and could distract these agents from extracting meaningful compositional patterns.

All the proposed experiments are then sent to A-Lab GPSS for testing (Figure \ref{fig:fig2}e). After synthesis, the newly analyzed experimental outcomes are added to the experimental dataset. The dataset is then clustered by composition and fed into the agents as texts (see Methods). At each iteration of the closed-loop workflow, two agents are invoked to carry out complementary modes of scientific reasoning: abductive reasoning via the abnormality-detection agent and inductive reasoning via the pattern-finding agent. The latter is then replaced by a BO-assisted pattern-finding agent when the experimental dataset is large enough. In the campaign detailed in the next section, the transition was made after the 289\textsuperscript{th} sample, a threshold chosen heuristically based on the amount of accumulated data. Working together, these agents form an explicit reasoning loop for autonomous experimental design that mirrors how human researchers operate, but with a larger number of samples than is feasible in standard laboratory workflows. 

\subsection{Self-driving discovery for lithium halide spinel ionic conductors}
We applied the self-driving A-Lab GPSS system to a synthesis campaign targeting lithium halide spinels with high ionic conductivity. Halide-based lithium ionic conductors have recently emerged as a promising material class for solid electrolytes in next-generation batteries, attributed to their high Li\textsuperscript{+} conductivities, good deformability, and electrochemical stability. \cite{li2020progress, tang2025halide} Within this landscape, spinel, with a general formula of \ce{Li_{2-$x$}M_{1±$y$}Cl_{4}}, stands out as a promising structure family to achieve high ionic conductivity\cite{liu2024li2fecl4, yang2025harnessing, wang2025cation, fu2025cost}. M represents one or multiple metal cations. $x$ denotes the degree of off-stoichiometry on the Li site. $y$ denotes the degree of off-stoichiometry on the metal site.

In this study, we focus on chloride lithium spinels, considering the broad accessibility and chemical diversity of commercially available metal chlorides. The chloride spinels have shown promising ionic conductivity, high stability, and tolerance to off-stoichiometry when doped with various metals. \cite{yang2025harnessing,rom2024expanding, jeon2024enhancing} Yet their vast compositional and processing design space remains sparsely mapped, largely because the experimental synthesis and property evaluation of these materials require a strictly air-free environment. This practical challenge can be tackled by designing a system like the A-Lab GPSS, which performs experiments in an air-free chamber and is further enhanced for increased efficiency by leveraging robotics and AI guidance.

In this experimental campaign, 352 samples with a target formula of \ce{Li_{2-$x$}M_{1±$y$}Cl_{4}} were synthesized via solid-state synthesis, characterized, and had their ionic conductivities measured within the A-Lab GPSS platform. The first 77 samples comprise initial attempts using experimental parameters designed by human researchers, combined with robotic experimentation. These 77 samples include a mix of monocations (single divalent cation), binary-cation compositions (isovalent/aliovalent substitution), random compositions, and a targeted study on the Mn\textsuperscript{II}-V\textsuperscript{III} spinel system. (Supplementary Note \ref{si-note:initial-human-test}, Supplementary Table \ref{si-table:initial-test}). This initial data set was designed to (i) validate the robustness of the experiment workflow for the target type of materials and (ii) generate seed data for the agentic workflow. After obtaining this initial test, the LLM agents took control of designing experimental parameters to continuously guide the closed-loop synthesis utilizing A-Lab GPSS. During this synthesis campaign, 12 additional samples in total were designed and submitted manually. The remaining 263 samples were proposed and executed in a self-driving manner. This campaign spanned 53 days from Nov. 03, 2025, to Dec. 26, 2025, utilizing a combination of fully automated robotic synthesis, semi-automated characterization and impedance measurement, and AI-driven decision-making to iterate on the experimental design.

\begin{figure}
    \centering
    \includegraphics[width=0.85\linewidth]{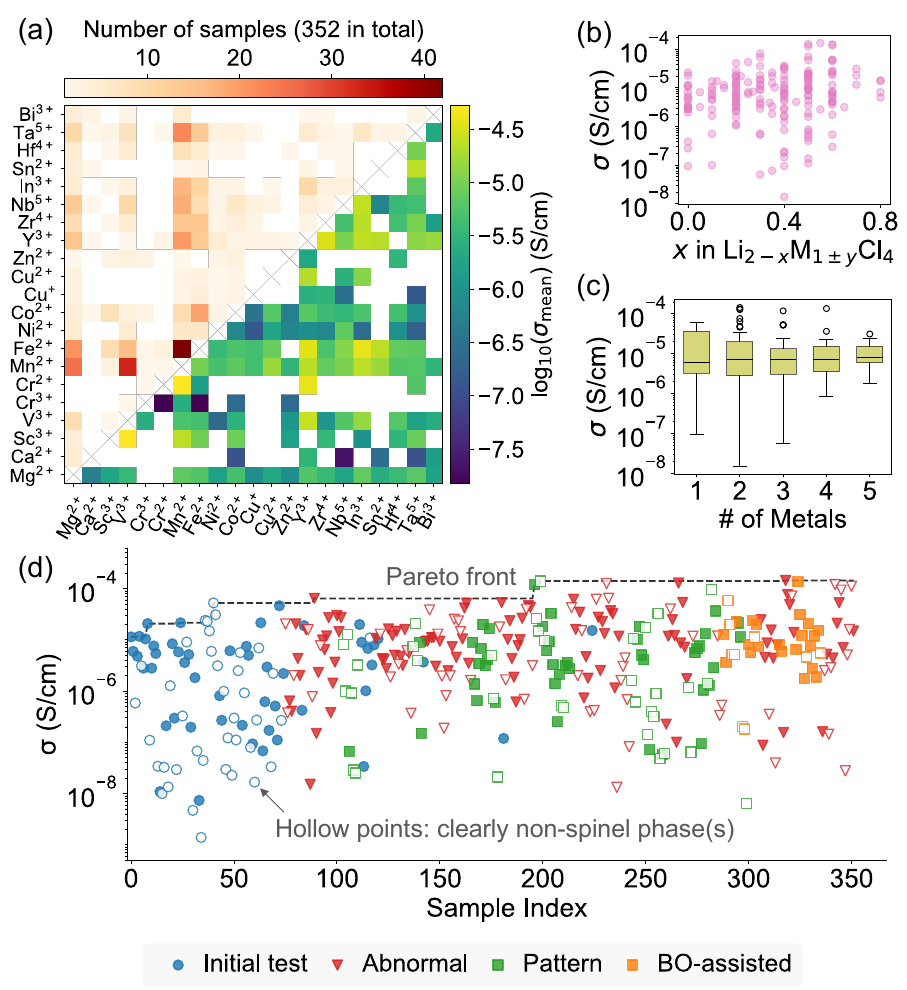}
    \caption[Statistical analysis of the experimental results.]{\textbf{Statistical analysis of the experimental results.}
    (a) Co-occurrence matrix of cations across the campaign. The upper triangle reports the number of samples containing each cation pair, while the lower triangle reports the mean ionic conductivity for that pair. 
    (b) Ionic conductivity versus lithium deficiency $x$ in the target composition of the synthesis, within formula \ce{Li_{2-$x$}M_{1±$y$}Cl_{4}}. M represents the metal species other than Li. $y$ denotes the degree of metal-site off-stoichiometry. Only samples with high spinel content (spinel phase weight fraction $>80\%$ and no clear unindexed peaks with relative intensity $>0.3$ normalized by the highest peak in the XRD pattern) are shown in this plot.
    (c) Boxplots of ionic conductivity grouped by the number of distinct metal species in the sample. Only high-spinel-content samples are shown. The center line denotes the median; the box edges correspond to the 25th and 75th percentiles. Circle markers indicate outliers beyond 1.5 times the interquartile range (IQR).
    (d) Temporal evolution of ionic conductivity across the campaign. Hollow markers indicate samples containing clearly identifiable non-spinel phases, either having a spinel phase weight fraction $<80\%$ or exhibiting a clear unindexed peak with relative intensity $>0.3$ compared to the most intense peak in the XRD pattern. The dashed line represents the running Pareto front of ionic conductivity.
}
    \label{fig:fig3}
\end{figure}

Figure \ref{fig:fig3} summarizes the results from the synthesis campaign. In total, 21 metal halide precursors were used, covering 19 metals and multiple oxidation states for some elements, as shown along the axes of Figure~\ref{fig:fig3}a. Given this large compositional space, we first quantify the diversity of the explored chemistry by counting the frequency with which each cation pair co-occurs in the same sample (upper triangle of Figure~\ref{fig:fig3}a). Across the campaign, which covered 352 samples, 124 distinct metal pairs were explored. Relative to the 171 possible metal pairs within this precursor set (Supplementary Note \ref{si-note:pw-calculation}), this corresponds to a pairwise coverage of 72\%. Meanwhile, a great number of trials were concentrated in specific compositional regions (darker colored cells in the upper triangle), particularly combinations of Fe$^{2+}$ or Mn$^{2+}$ with a higher-valent cation such as Y$^{3+}$, Zr$^{4+}$, Nb$^{5+}$, and Ta$^{5+}$. Notably, these same regions exhibit higher mean ionic conductivities (brighter cells in the lower triangle), suggesting that the agents tend to explore these chemically favorable subspaces as evidence accumulates.

We next examine the relationship between lithium deficiency and ionic conductivity, motivated by recent reports that Li vacancies can facilitate Li-ion transport via vacancy-mediated mechanisms in halide spinel frameworks.\cite{rom2024expanding, yang2025harnessing, zhou2025vacancy} Figure~\ref{fig:fig3}b plots each high-spinel-content sample as a point, with the $x$-axis representing the targeted lithium deficiency $x$ in \ce{Li_{2-$x$}M_{1±$y$}Cl_{4}}. The upper envelope of ionic conductivity increases with increasing Li deficiency up to a Li deficiency level of 0.5-0.6, beyond which the maximum achievable conductivity declines. Although ionic conductivities can vary with different chemical compositions at the same level of Li deficiency, the observed trend suggests that ionic conductivity correlates with lithium deficiency in the target compositions. This non-monotonic behavior indicates the presence of an optimal vacancy concentration for achieving high ionic conductivity. A moderate degree of Li deficiency enhances Li-ion diffusion within the structure, whereas excessive Li deficiency may reduce the effective carrier density in the percolating Li-ion sublattice or reduce Li mobility by Li vacancy ordering, thereby constraining long-range ionic transport. 

We further probe cation-entropy effects by grouping the high-spinel-content samples according to the number of distinct metal species in each sample (Figure~\ref{fig:fig3}c). Notably, increasing the number of cations does not systematically raise the maximum observed conductivity. Instead, the distribution shifts upward: as the number of metal species increases from two to four, the conductivity range narrows, and the median converges toward a higher value. For five-metal compositions, the conductivity span appears even smaller, although this is also related to the limited number of samples explored in this group.
While these trends should be interpreted with caution due to incomplete sampling, this behavior is consistent with a disorder-assisted transport picture in which multi-cation configurations introduce local structural disorder, creating a percolating pathway with lower effective migration barriers.\cite{zeng2022high}

Figure~\ref{fig:fig3}d tracks the temporal evolution of phase purity and ionic conductivity throughout the campaign. Samples are ordered by submission time, and marker shapes indicate provenance: initial human-driven samples (``Initial test''), samples selected by the abnormality-detection agent (``Abnormal''), the pattern-finding agent (``Pattern''), and the BO-assisted pattern-finding agent (``BO-assisted''). Hollow markers denote samples with clearly identifiable non-spinel signatures in XRD (weight fraction of spinel phase $<80\%$ or a prominent unindexed peak with relative intensity exceeding 30\% of the maximum XRD peak). The dashed line traces the running Pareto frontier, highlighting the best conductivity achieved at each point in time and progressing as more samples are tested.

The success rate of the agent-driven workflow increases over time. Here, a sample is defined as successful if it exhibits both good ionic conductivity ($> 0.05$ mS/cm) and high spinel purity, where high spinel purity is defined as a spinel phase weight fraction greater than 80\% with no clear unindexed peaks whose relative intensity exceeds 0.3 relative to the most intense XRD peak. Using this definition, a clear upward trend in this success rate is observed (Supplementary Figure~\ref{si-fig:success}). In the initial human-guided test set, none of the samples met these criteria. After the AI agents took over, the success rate among the first 75 agent-proposed samples was 1.33\%, whereas the success rate among the final 75 samples increased to 5.33\%. This approximately fourfold improvement in hit rate further indicates that the closed-loop, agent-driven workflow progressively concentrates experiments in chemically favorable regions of the search space.

Among all samples, a few promising chemical spaces stand out, exhibiting ionic conductivities exceeding 0.1 mS/cm. The representative compositions are \ce{Li_{1.45}Mn_{0.45}Sc_{0.55}Cl4} (0.144 mS/cm, Supplementary Note \ref{si-note:li-mn-sc-cl}), \ce{Li_{1.5}Mn_{0.25}Fe_{0.25}Y_{0.25}In_{0.25}Cl4} (0.136 mS/cm, Supplementary Note \ref{si-note:li-fe-mn-in-y-cl}). Interestingly, for the first composition, a closely related Li-Mn-Sc-Cl composition (\ce{Li_{1.5}Mn_{0.5}Sc_{0.5}Cl4}) was included in the initial screening but showed an ionic conductivity of only 0.046 mS/cm. This underperformance can be attributed to incomplete reaction between precursors at the initially applied synthesis temperature (450 $^\circ$C), as indicated by minor residual peaks in the XRD pattern, likely due to the presence of \ce{ScCl3}. After the abnormality-detection agent identified this bottleneck, it progressively increased the synthesis temperature from 500 $^\circ$C to 550 $^\circ$C, thereby enabling a more complete reaction and ultimately delivering higher conductivity.
The second outstanding composition, from the Li-Mn-Fe-Y-In-Cl system, emerged as a promising ionic conductor during the middle stage of exploration (first identified as \ce{Li_{1.5}Mn_{0.3}Fe_{0.2}Y_{0.25}In_{0.25}Cl4} (0.140 mS/cm)). However, XRD measurements consistently showed the presence of a secondary \ce{Li3YCl6} or \ce{Li3InCl6} phase (around 15 - 30 wt\%), which is flagged by the abnormality-detection agent as an anomalous outcome. Inspection of the agent’s reasoning traces shows that it hypothesizes that the impurity phase can be partially responsible for the elevated conductivity, given that \ce{Li3YCl6} and \ce{Li3InCl6} are known fast ionic conductors.\cite{schlem2021insights, li2019back} Because the agents were explicitly instructed to prioritize compositions that can form spinel structures, they submitted multiple follow-up experiments to eliminate the \ce{Li3YCl6}/\ce{Li3InCl6} phase. However, even after 352 experiments, the agents were unable to completely eliminate these secondary phases while preserving high conductivity, suggesting that the high ionic conductivity arises from the close-packed \ce{Li3YCl6}/\ce{Li3InCl6} phase.
Overall, these cases highlight a central challenge in autonomous materials discovery: improving target properties while disentangling (and minimizing) the influence of persistent secondary phases through careful synthesis control and phase-purity optimization. In this campaign, even with only one property to optimize, many efforts are still needed to find a material with good ionic conductivity while retaining the spinel structure. 

\subsection{Behavior analysis of the agents}
The agents display human-like behavior in their reasoning process. They examine prior experimental results, apply general chemical intuition, and propose targeted follow-up experiments designed to isolate causal factors. For example, the initial test includes ionic conductivities for the nominal endmembers \ce{Li2MnCl4} ($5.59\times10^{-4}$~mS/cm) and \ce{Li2FeCl4} ($1.14\times10^{-2}$~mS/cm). When a doubled ionic conductivity was observed in  \ce{Li_{1.8}Fe_{0.8}In_{0.2}Cl4} ($2.72\times10^{-2}$~mS/cm, compared to \ce{Li2FeCl4}), the abnormality-detection agent asked a question: ``does high $\sigma$ in \ce{Li_{1.8}Fe_{0.8}In_{0.2}Cl4} arise from intrinsic vacancy/disorder (y$\approx$0.2) or undetected Li-In-Cl grain-boundary phases (hinted by the presence of LiCl phase in the sample)?'' To disentangle these effects, the agent designed \ce{Li_{1.8}Fe_{1.1}Cl4}, removing indium while preserving lithium deficiency. The resulting conductivity increased further to $4.30\times10^{-2}$~mS/cm (nearly fourfold relative to \ce{Li2FeCl4}). This outcome was later generalized by the pattern-finding agent, as revealed by its reasoning trace, into a transferable design rule: introducing excess metal content into the spinel framework can be an effective route to create Li vacancies and increase ion transport. Noticing the chemical similarity between \ce{Fe^{2+}} and \ce{Mn^{2+}}, the pattern-finding agent applied the same strategy to \ce{Li2MnCl4} and designed \ce{Li_{1.8}Mn_{1.1}Cl4}, ultimately achieving a leap to $1.75\times10^{-2}$~mS/cm (over an order-of-magnitude improvement of \ce{Li2MnCl4}'s $5.59\times10^{-4}$~mS/cm). The details about the reasoning trajectory are shown in Supplementary Note \ref{si-note:li-mn-cl}. Although a similar composition (\ce{Li_{1.6}Mn_{1.2}Cl4}) was reported previously \cite{kanno1984ionic}, and this chemical system was systematically studied very recently \cite{tao2026spinel} (published after this campaign), the AI agents independently rediscovered this sample through step-by-step hypothesis formation grounded in the accumulating dataset.

\begin{figure}
    \centering
    \includegraphics[width=\linewidth]{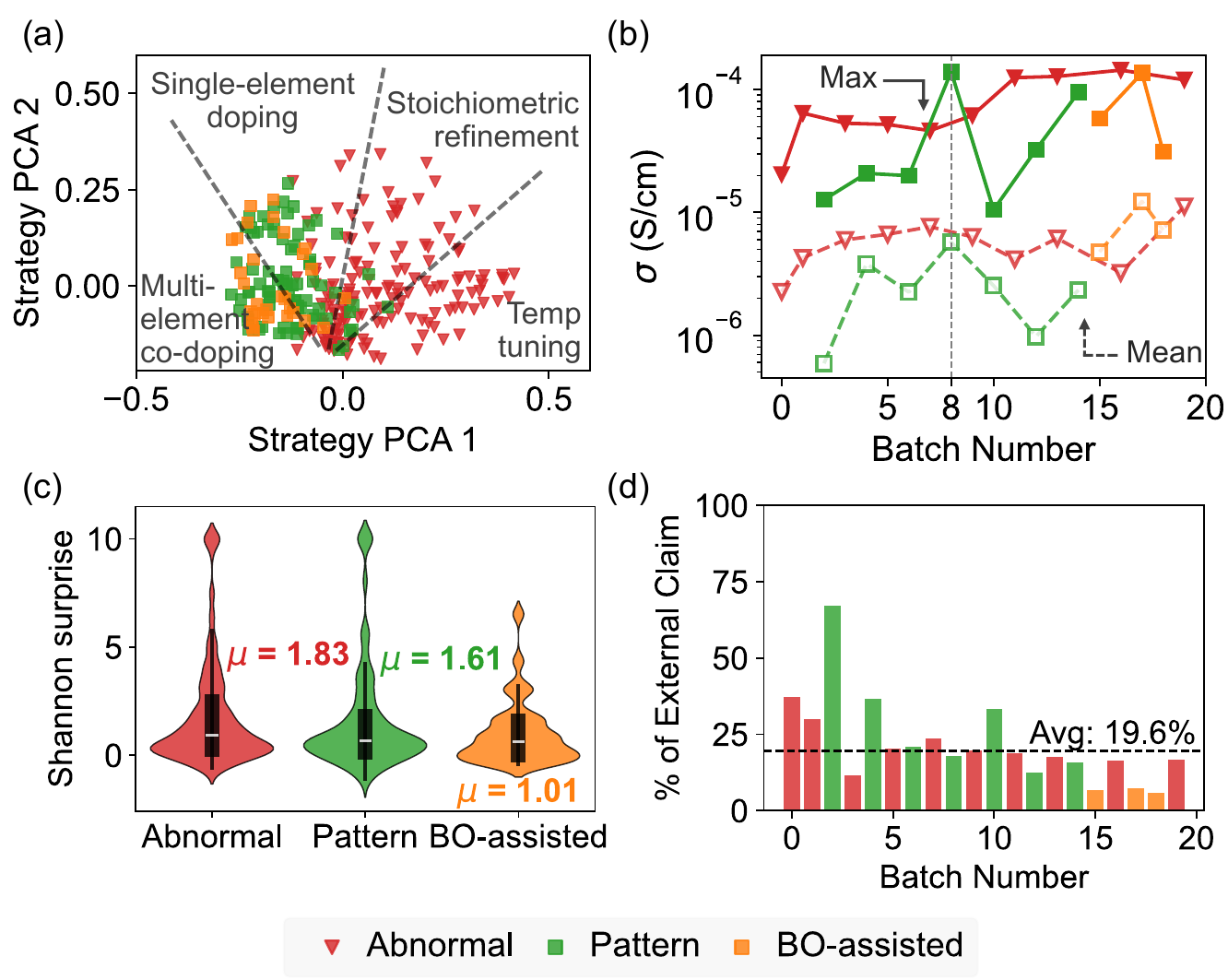}
    \caption[Behavior analysis of experimental design agents.]{\textbf{Behavior analysis of experimental design agents.}
(a) Principal component analysis (PCA) of strategy embeddings derived from agents' experiment-design justification. Four dominant strategy clusters emerge and are labeled accordingly: temperature tuning (shortened to ``Temp tuning'' in the figure), stoichiometric refinement, single-element doping, and multi-element co-doping.
(b) The max (solid line) and mean (dashed line) values of all samples' ionic conductivities over batches.
(c) Shannon surprise of samples proposed by each agent; the mean surprise ($\mu$) is annotated beside each violin plot. Surprise values are clipped at 10 to mitigate numerical instabilities arising from the logarithmic operation.
(d) Ratio of external claims (statements not directly grounded in prior experimental data) versus batch number.
Across panels, color indicates sample provenance: red (abnormality-detection agent), green (pattern-finding agent), and orange (BO-assisted pattern-finding agent).
}
    \label{fig:fig4}
\end{figure}

Figure~\ref{fig:fig4} shows quantitative analysis of the agents' behaviors. We first analyze how different agents propose new experiments given the accumulated data. For each proposal, the agent generated a textual justification describing both its motivation and the specific modification introduced. We then use an LLM to distill each justification into a short strategy statement (e.g., ``increase the temperature to 450 $^\circ$C'' or ``introduce Mn as a new dopant''). Text embeddings of these strategy statements are computed using \texttt{gemini-text-embedding-001}, followed by principal component analysis (PCA) for visualization. As shown in Figure~\ref{fig:fig4}a, each point represents the strategy used to design a single new experiment and is color-coded by the agent responsible for that proposal. The abnormality-detection agent and the pattern-finding agents occupy distinct regions of embedding space, indicating systematic differences in their design policies. Manual inspection over the extracted strategy texts identifies four dominant strategies: (i) temperature tuning, (ii) stoichiometric refinement, (iii) single-element doping, and (iv) multi-element co-doping. The abnormality-detection agent preferentially performed modifications that preserve elemental composition, whereas the pattern-finding agents more frequently introduce new elements, expanding chemical space. This divergence reflects the different reasoning modes specified in their prompts: for the abnormality-detection agent, control experiments help interpret anomalous samples, whereas for the pattern-finding agents, inferred patterns more often involve chemical substitutions and their effects on phase purity and ionic conductivity.

The different roles of the agents are further reflected in the batchwise evolution of ionic conductivity (Figure~\ref{fig:fig4}b). Because a primary objective of the campaign was to maximize the ionic conductivity of the synthesized samples, this metric provides a direct readout of how each agent pursued the optimization goal. Before batch \#8, no sample's ionic conductivity exceeded 0.1~mS/cm ($1\times10^{-4}$ S/cm), with the abnormality-detection agent typically reaching batch maxima between 0.01 and 0.08 mS/cm. In the batch \#8, the pattern-finding agent identified \ce{Li_{1.5}Mn_{0.25}Fe_{0.25}Y_{0.25}In_{0.25}Cl4} with ionic conductivity \textgreater 0.1 mS/cm. Following this discovery, the abnormality-detection agent rapidly converged toward batch maxima in the 0.1 mS/cm regime, suggesting that it prioritized the newly discovered high-conductivity region, whereas the pattern-finding agent and its BO-assisted variant continued to exhibit larger batch-to-batch fluctuations. The batchwise mean ionic conductivities (dashed lines) also support this exploration–exploitation pattern: the abnormality-detection agent maintains a higher, more stable average, whereas the pattern-finding agent exhibits substantially greater variability. It is worth noting that the BO-assisted pattern-finding agent also achieves a relatively high mean conductivity, likely because Bayesian optimization constrains the initial candidate pool to those with good predicted ionic conductivity, thereby biasing selection toward more promising compositions.

We next quantify exploration using Shannon surprise, defined as the negative log-likelihood of the observed outcome under a Gaussian-process (GP) model trained without the new experimental observation. For each sample, a GP model was fit using all preceding data to estimate the likelihood of the measured ionic conductivity conditioned on the previously explored samples (Supplementary Note \ref{si-note:shannon-surprise}). Shannon surprise thus provides a measure of how unexpected each experimental outcome is relative to prior observations, with higher values indicating experiments that contribute greater new information to the model/whole dataset.
As shown in Figure~\ref{fig:fig4}c, the abnormality-detection agent achieves the highest mean surprise (1.83), followed by the pattern-finding agent (1.61), while the BO-assisted agent exhibits lower surprise (1.01), consistent with its value-uncertainty trade-off. The high surprise values produced by the abnormality-detection agent may reflect a highly non-smooth target-property landscape, in which minor changes in composition or synthesis conditions can lead to substantial variations in ionic conductivity, potentially driven by the formation of secondary phases or precursor evaporation at elevated synthesis temperatures.
These results indicate that, although the abnormality-detection and pattern-finding agents adopt different design strategies, both are capable of identifying undersampled regions of the search space and proposing experiments that yield informative, high-surprise outcomes.

During experimental design, the agents can ground their claims in two sources of knowledge: (1) results directly from previously explored samples in the campaign, provided in the prompt as in-context experimental evidence (dataset-referenced), and (2) external knowledge acquired from the previous literature in the training corpus (external-referenced). In addition to each proposed experiment, the agents also produce a textual justification that exposes their reasoning process. The knowledge invoked in these justifications can be decomposed into individual claims, each representing a specific piece of evidence or background knowledge used in the design decision. To analyze how agents reference knowledge during the campaign, we extract the claims in the justification and classify them as dataset-referenced or external-referenced using an independent LLM workflow (Supplementary Note \ref{si-note:claim-extraction}). Among all the claims, 19.6\% of claims require knowledge from previous scientific literature. As shown in Figure~\ref{fig:fig4}d, the fraction of external-referenced claims decreases over successive batches ($\rho_\text{Spearman} = -0.7489$, $p = 1.42 \times 10^{-4}$), indicating decreased reliance on external-referenced knowledge as the dataset expands. \cite{hauke2011comparison}

\section{Discussion}
We developed A-Lab GPSS, an automated platform integrating air-sensitive solid-state synthesis with semi-automated characterization. We demonstrated its LLM-agent-driven operation towards the exploration and optimization of halide spinel conductors. By explicitly structuring autonomous experimental design into abductive (abnormality-detecting) and inductive (pattern-finding) reasoning modes, we enable LLM agents to play complementary, synergistic roles in designing experiments. Across a 352-sample campaign involving 19 metals (21 cation species), A-Lab GPSS increased the success rate for identifying materials with both good ionic conductivity and high spinel purity from 1.33\% (first 75 samples) to 5.33\% (last 75 samples), while achieving 72\% pairwise coverage of this broad chemical space.

Furthermore, by making AI agents’ reasoning processes observable, we identified that the abductive and inductive reasoning agents tend to employ distinct yet complementary strategies for proposing new experiments: the abductive reasoning agent focuses on anomalous observations within already explored regions, enabling finer local re-exploration, whereas the inductive reasoning agents more often expand the search into broader and previously unvisited chemical space. Together, they form a synergistic discovery framework that balances re-exploration of known chemical space with broad exploration of new compositional regions. 

At the same time, this study also highlights an important limitation of the current platform. Sample characterization was limited to powder XRD and EIS measurement, whereas more in-depth structural characterization is often required to resolve fine structural features and establish causal links between structure and property. In the absence of such information, the agents must sometimes infer underlying mechanisms from limited evidence, which can reduce the efficiency and reliability of experimental design. For example, the agents frequently invoked cation-ordering effects to explain differences in ionic conductivity, yet these hypotheses could not be validated because the current system lacks characterization tools capable of directly probing such structural details. This limitation points to two major directions for future development: first, integrating broader multimodal characterization capabilities into the laboratory platform to provide stronger evidence for agent reasoning; and second, equipping agentic AI systems with calibrated uncertainty estimates so they can quantify confidence in their hypotheses and defer to human judgment when uncertain. Progress in both directions will be essential to building more reliable and efficient self-driving laboratories to accelerate high-throughput materials discovery.

\section{Methods}
\subsection{Automated solid-state synthesis in A-Lab GPSS}
All syntheses are performed in a \ce{N2}-filled glovebox, where the oxygen concentration is maintained below 0.1 ppm and the water concentration below 10 ppm. The glovebox is equipped with an air-conditioning system set to 24$^\circ$C ($\pm$1$^\circ$C) to dissipate heat generated by the furnaces. The synthesis follows a conventional solid-state route. All the chemicals used in this study are listed in Supplementary Table \ref{si-tab:precursors}. Two robot arms (Universal Robots UR5e) are used to transfer samples between workstations. A linear rail system bridges the two arms, enabling sample handoff without requiring the robots to be positioned in close proximity. 

Reactions are carried out in pure nickel crucibles to minimize chemical reaction between the crucible and halide precursors at elevated temperatures. Each crucible is sealed with a nickel-plated stainless-steel cap, which is mechanically clamped to reduce materials evaporation during heating. A clean crucible is assigned to each sample at the time of submission. The synthesis begins with a robotic arm removing the crucible cap, followed by dispensing four nickel milling balls into the crucible. Powder precursors are then dispensed sequentially using an automated Mettler Toledo balance (XPR305D5Q), with a dispensing tolerance of $\pm$0.1\%. After dispensing, the crucible is resealed with the metal cap.
The sealed crucible is transferred to a dual asymmetric centrifuge (Smart DAC400, Hauschild) for mixing, operated at 1000 rpm for 5 min, followed by 1500 rpm for an additional 5 min. After mixing, up to six crucibles are placed onto a heating rack. Using a custom CNC-machined stainless-steel furnace handle, the robot transfers the rack into a box furnace. Two furnaces (Thermo Fisher F48055-60) are operated in parallel to simultaneously heat two batches of samples.
Samples are heated to the target temperature at a ramp rate of 2 $^\circ$C/min, held for 12 h, and then allowed to cool naturally to 100 $^\circ$C before removal from the furnace. After heating, the crucible cap is removed, and two additional nickel milling balls are added. The crucible is then transferred to a vertical shaker, where the balls are shaken within it to grind the synthesized material into a fine powder. The resulting powder is poured into a plastic vial by the robot arm.
For XRD sample preparation, a plastic cap with a central opening covered by a stainless-steel sieve is attached to the vial. The vial is placed onto a rotary gripper equipped with an embedded vibrational motor. During sample dispensing, the gripper inverts the vial over the center of the XRD sample holder while the vibrational motor actuates, dispensing a controlled amount of powder onto the holder. After dispensing, the sieved cap is replaced with a standard plastic cap to seal the vial for sample storage. Details about consumables used in A-Lab GPSS are described in Supplementary Note \ref{si-note:consumbles}.
The workflow management framework was developed in-house as part of AlabOS \cite{fei2024alabos}. All the control code and workflow definitions are available in the Code Availability section.

\subsection{Semi-automated characterization workflow}
After completion of the synthesis workflow, several manual steps are performed for characterization. The sample powders on XRD sample holders are covered with a thin Kapton film inside the glovebox to prevent air exposure. The holders are then loaded into a benchtop X-ray diffractometer (Malvern Panalytical Aeris Mineral Edition). Each XRD scan is collected from 10$^\circ$ to 100$^\circ$ (2\texttheta) with a total acquisition time of 8 minutes. The diffraction XRD patterns are automatically parsed and uploaded to AlabOS’s central database, where all information of a sample is stored.

To measure ionic conductivity, electrochemical impedance spectroscopy (EIS) is utilized. The powder samples generated from the automated synthesis workflow are loaded into a customized measurement cell with two stainless-steel rods functioning as electrodes (0.256 inch in diameter) and a nylon spacer to confine the powder. The cell is placed in an automated hydraulic press (MSE Supplies PR0342), where 2 tons of pressure is applied for 30 seconds to form a dense pellet. The top pressing rod and base plate are connected through a glovebox feedthrough to an electrochemical workstation (BioLogic SP 300). The wiring resistance (1.5-2 \textOmega) is negligible compared to the sample resistance.
Impedance spectra are collected over a frequency range of 7 MHz to 1 Hz, with two repeated measurements per sample. After measurement, the pellet thickness is measured using a digital caliper. All characterization data, including the full impedance spectrum and sample thickness, are automatically uploaded to AlabOS central database for downstream analysis.

\subsection{Automatic analysis of XRD patterns}
Each sample's XRD pattern is analyzed using Dara \cite{fei2026dara} for phase identification and basic Rietveld refinement. The reference phases are downloaded from the ICSD database and used for phase identification. A custom spinel structure with the target composition is also generated and included in the phase search as the structure model for the sample's targeted spinel.
The maximum phase search depth is set to 4. The stopping criteria for the search require a positive improvement in $R_{pb}$ (\textgreater 0) to continue. For all ICSD reference phases, we allow a preferred orientation of the fourth-order spherical harmonic function (SPHAR4) and a 5\% lattice parameter shift. The allowed peak broadening parameter B\textsubscript{1} is set to be 0-0.08 for the spinel phase, whereas for all other phases, it is set to be 0-0.01. The only exception was in the first two batches, where the B\textsubscript{1} range was set to 0-0.08 for all phases (Supplementary Note~\ref{si-note:changes}). Other parameters are set to their default values. After the search, the top phase identification result with the lowest $R_{wp}$ is taken as the identified phases, with the missing and extra peaks marked in the analyzed result.

After the synthesis campaign, we also manually re-inspected all the XRD patterns; the results are in Supplementary Data. While the automated analysis successfully identifies the major phases across all patterns, a few specific discrepancies are noted: (1) overfitting with multiple spinel phases to one peak, which is a result of an occasional strong preferred orientation effect caused primarily by non-ideal sample preparation; (2) underfitting with minor impurity phases with low intensity or peak overlapping with major phases. Although these issues can occasionally hinder an agent's ability to extract accurate phase fractions, they are mitigated by the fact that the agents have access not only to the final phase fractions but also to the quality of fit (weighted profile residual $R_{wp}$) and the space group information for each phase.

\subsection{Automatic analysis of EIS spectra}
The EIS spectra are fitted to a custom circuit using the \texttt{impedance.py} package \cite{Murbach_impedance_py_A_Python_2020}. A fixed circuit of ``CPE\textsubscript{1}--(CPE\textsubscript{2}, R\textsubscript{1})'' is used for all the samples. An evolutionary algorithm in SciPy \cite{2020SciPy-NMeth} is used to find the optimal fit for the circuit parameters. In some cases, the EIS spectra do not fit the circuit model well, which may be due to other electrochemical behavior of the sample. The algorithm flags all the spectra with an RMSE \textgreater 1.0 for human inspection. The ionic conductivity is calculated with the measured impedance, sample thickness, and the area of the stainless-steel rod. The calculated ionic conductivity is provided to the agents as a numerical value. During the campaign, the ionic conductivity values for 13 samples were updated following issues identified in the analysis workflow, as discussed in Supplementary Note~\ref{si-note:changes} and summarized in Supplementary Table~\ref{si-tab:changes}.

\subsection{Agent workflows for experimental design}
The LLM agents are implemented with LiteLLM; unless otherwise stated, OpenAI GPT-5 with high reasoning effort is used as the base model. Different prompts are provided to drive the agents' abductive/inductive reasoning; these are available in the Code Availability section.

All prior experimental data are provided to the agents as a structured JSON string. Each entry includes the target composition, synthesis temperature, XRD analysis results, ionic conductivity, and a short experiment note. The note is generated by a separate LLM after synthesis and characterization, summarizing the experimental rationale and key conclusions (discussed later in this section). For XRD results, we include not only the identified phases and their weight fractions, but also the corresponding space groups and the refinement quality metric ($R_{wp}$), enabling the agents to reason with richer structural information.

During the campaign, experimental records are incorporated directly into the user prompt as in-context learning. Because this approach is constrained by the finite context window of the LLM, we implement a clustering strategy to ensure scalability. Motivated by the heuristic that compositionally similar samples provide the most informative comparisons, we apply a modified hierarchical clustering algorithm to partition the dataset. The algorithm recursively performs agglomerative clustering with progressively tighter distance thresholds until all clusters contain fewer than 80 samples (approximately 30,000 tokens). To preserve cross-cluster continuity, each cluster additionally includes 10 overlapping samples from neighboring clusters. The agentic workflow is then executed independently within each cluster.

For the abnormality-detection agent, experimental design proceeds through two sequential LLM calls. In the first call, the agent analyzes the existing experimental dataset to identify abnormal samples. For each identified abnormal sample, the model outputs: (1) the sample composition and synthesis condition, (2) a justification explaining why the sample is considered abnormal, and (3) a set of relevant reference samples for comparison.
In the second call, the agent formulates a hypothesis to explain the identified abnormality and proposes a follow-up experiment. The output includes: (1) a hypothesis that potentially accounts for the observed anomaly, (2) a new experimental design (composition and synthesis temperature) intended to test this hypothesis, and (3) a justification explaining how the proposed experiment discriminates among competing explanations.
The pattern-finding agent issues only one LLM call. The agent analyzes accumulated data to infer emerging trends and directly outputs a list of new candidate samples, each accompanied by a justification specifying (1) the inferred pattern and (2) the rationale for why the proposed composition is expected to validate or extend that pattern. The synthesis temperature for each composition is determined using a modified Tammann’s rule, where three-quarters of the average melting point of the precursor salts is used as the target synthesis temperature.
The BO-assisted pattern-finding agent follows a similar structure, but incorporates an additional Bayesian optimization (BO) step prior to LLM reasoning. A Gaussian-process-based BO model is first trained on the existing dataset to generate a ranked list of promising candidate compositions (see Supplementary Note \ref{si-note: bo-training}). The LLM then selects and justifies experiments from this candidate pool.
Before any LLM-designed experiment is executed in A-Lab GPSS, it is evaluated against a set of manually defined feasibility and safety constraints to ensure compatibility with the hardware and experimental protocols (see Supplementary Note \ref{si-note:sanity-check}). Only experiments that satisfy these pre-screening rules are submitted for autonomous execution.

After each experiment, another LLM call is made to summarize the outcome and explicitly compare the results to the original design justification. Medium reasoning effort is used for this task to increase efficiency. This structured reflection serves as a memory mechanism, allowing the agents to track prior hypotheses, evaluate their validity, and inform subsequent decision-making.

\section{Data and code availability}
\begin{itemize}
    \item The code to control A-Lab GPSS and run LLM agents is available at \url{https://github.com/CederGroupHub/alab_gpss_public}.
    \item The experimental dataset of halide spinel samples is available at \url{https://doi.org/10.5281/zenodo.19396297}.
    \item The video of A-Lab GPSS is available at \url{https://youtu.be/9j-3aIf02jE}. 
\end{itemize}

\section{Acknowledgement}
This work was primarily funded by the Data Driven Synthesis Science Program (D2S2) of the U.S. Department of Energy, Office of Science, Basic Energy Sciences, under Contract No. DE-AC02-05CH11231. Building of the GPSS experimental setup was funded in part by the Energy Storage Research Alliance (ESRA), an Energy Innovation Hub funded by the U.S. Department of Energy, Office of Science, Basic Energy Sciences, under Contract No. DE-AC02-06CH11357. Specific materials and electrochemical testing were performed under the Advanced Battery Material Research (BMR) Program under Contract No. DE-AC02-05CH11231.
B.R. acknowledges support from the Kavli ENSI Graduate Student Fellowship.
S.W. acknowledges partial funding of the Jane Lewis Fellowship at UC Berkeley.
This research uses the CBorg AI platform and resources provided by the IT Division at the Lawrence Berkeley National Laboratory (Supported by the Director, Office of Science, Office of Basic Energy Sciences, of the U.S. Department of Energy under Contract No. DE-AC02-05CH11231).

The authors thank Emory Chan (LBNL), Michael Whittaker (LBNL), Deepak Rawat (LBNL), Yizhou Zhu (Westlake University), and Shijing Sun (University of Cambridge) for helpful discussions and valuable input on this project.

\section{Author Contribution}
Y.F.: conceptualization, methodology, software, writing - original draft, writing - review and editing. 
B.R., J.W., and D.M.: conceptualization, methodology, software, writing - review and editing.
X.Y., C.L., and S.W.: conceptualization, methodology, writing - review and editing.
X.H.: software, writing - review and editing.
Y.Z.: supervision, methodology, project administration, writing - review and editing.
G.C.: funding acquisition, resources, supervision, methodology, project administration, writing - review and editing.

\section{Declaration of interests}
G.C. is a member of Joule’s advisory board.

\section{Declaration of generative AI and AI-assisted technologies}
During the preparation of this work, the author(s) used OpenAI GPT-5 in order to refine
the writing. After using this tool/service, the author(s) reviewed and edited the content as needed and take(s) full responsibility for the content of the published article.

\bibliography{ref}

\clearpage

\ifarXiv
    \foreach \x in {1,...,\numbersupplementpages}
    {
        \includepdf[pages={\x}, fitpaper=true]{\supplementfilename}
    }
\fi

\end{document}


\title{\textbf{Supplementary Information for ``Agentic LLM Reasoning in a Self-Driving Laboratory for Air-Sensitive Lithium Halide Spinel Conductors''}}

\author[1,2]{Yuxing Fei\,\orcidlink{0000-0002-1225-2083}}
\author[1,2,5]{Bernardus Rendy\,\orcidlink{0000-0001-8309-6279}}
\author[1,2]{Xiaochen Yang\,\orcidlink{0000-0002-8359-5630}}
\author[1,3]{Junhee Woo\,\orcidlink{0009-0002-1869-2861}}
\author[1,2]{Xu Huang\,\orcidlink{0009-0002-2260-5150}}
\author[2]{Chang Li\,\orcidlink{0000-0001-5420-3856}}
\author[1,2]{Shilong Wang\,\orcidlink{0009-0004-8504-5802}}
\author[2]{David Milsted\,\orcidlink{0000-0003-0415-910X}}
\author[2,4,$\dagger$]{Yan Zeng\,\orcidlink{0000-0002-5831-1210}}
\author[1,2,5,$\ddagger$]{Gerbrand Ceder\,\orcidlink{0000-0001-9275-3605}}
\affil[1]{Department of Materials Science \& Engineering, University of California, Berkeley, Berkeley, California 94720, United States}

\affil[2]{Materials Sciences Division, Lawrence Berkeley National Laboratory, Berkeley, California 94720, United States}

\affil[3]{Department of Materials Science and Engineering, Korea Advanced Institute of Science and Technology, Daejeon 34141, Republic of Korea}

\affil[4]{Department of Chemistry and Biochemistry, Florida State University, Tallahassee, Florida 32306, United States}

\affil[5]{Energy Storage Research Alliance, Argonne National Laboratory, Lemont, Illinois 60439, United States}
\affil[$\dagger$]{Correspondence: zeng@chem.fsu.edu}
\affil[$\ddagger$]{Correspondence: gceder@berkeley.edu}
\date{}

\maketitle
\setstretch{1.2} 
\newpage
\section*{Supplementary information}
\textbf{Document S1.} Notes S1--S11, Table S1--S3, Figures S1--S4, and supplementary references

\noindent \textbf{Video S1.} A-Lab GPSS synthesis workflow
\newpage
\section{Consumables in A-Lab GPSS}
\label{si-note:consumbles}
\subsection*{Nickel crucible and cap}
\begin{figure}[H]
    \centering
    \includegraphics[width=0.7\linewidth]{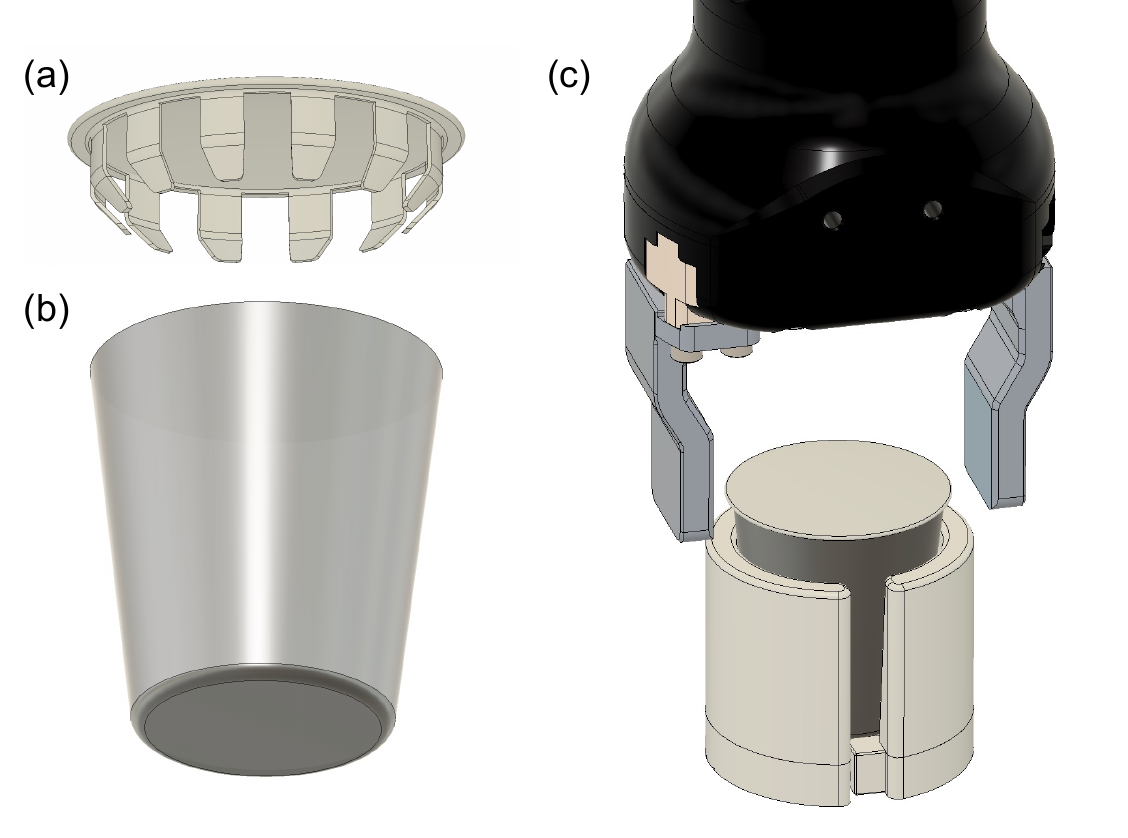}
    \caption{The design of the crucible and cap. (a,b) The 3D model of the crucible and the cap. (c) Illustration of a robot arm pulling the cap off the crucible, with the crucible held by a stationary gripper.}
    \label{si-fig:crucible-cap}
\end{figure}

Both the crucible and the cap used in A-Lab GPSS are commercially available, unmodified (Supplementary Figure \ref{si-fig:crucible-cap}a-b). The crucible is purchased from United Scientific Supplies, with SKU\# NCR025. The Nickel-plated cap is repurposed from the steel hole plug from Keystone Electronics with SKU\# 7610. The cap fits into the crucible and seals it by friction. When removing the cap from the crucible, a stationary gripper on the robot arm grasps the crucible and pulls the cap off from above to open it. (Supplementary Figure \ref{si-fig:crucible-cap}c)

After each reaction, the metal cap is disposed of and will not be reused. The crucible needs to be manually cleaned outside the glovebox before reuse in the next experiment. To clean the crucible, the residual powders are first dumped, then rinsed with tap water. In some cases, the metal chlorides hydrolyze to hydroxides, and a steel scrubber is used to remove the hydroxides that stick to the crucible wall. After cleaning, the crucible is completely dried in an oven at 70 $^\circ$C before being sent back to the glovebox for the next experiment.

\subsection*{Sample storage vial and tracking code}
The sample storage vial is purchased from Parkway Plastic (SKU\# A0430100PPC \& C043C4SPTSW). After the experiment, the samples are stored in the vial. For each sample, a tracking label is generated and manually attached to each jar before being removed from the glovebox for future reference. An example label is shown in Supplementary Figure \ref{si-fig:tracking-label}.

\subsection*{XRD sample holder}
\begin{figure}[H]
    \centering
    \includegraphics[width=0.8\linewidth]{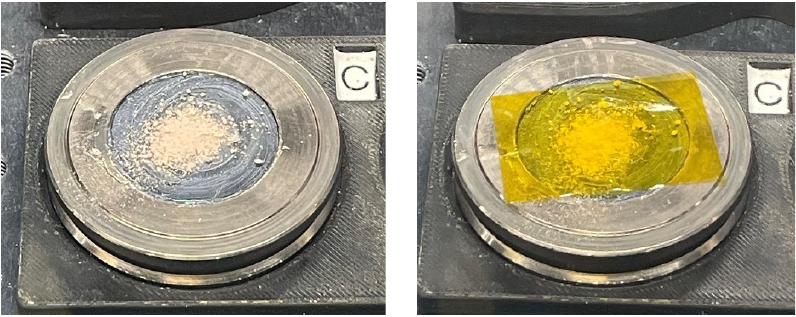}
    \caption{The XRD sample holder before and after applying the Kapton film.}
    \label{si-fig:xrd}
\end{figure}

The XRD sample holder is purchased from Malvern Panalytical (Catalog Nos. 9430 018 19321 and 9430 018 13321). Before usage, a thin layer of vacuum grease is applied to the zero-background to later stick the Kapton film. After the XRD samples are prepared, a Kapton film (McMaster Carr, SKU\# 2271K41) is applied to cover the XRD sample to protect it from the air during measurement (Supplementary Figure \ref{si-fig:xrd}). After the measurement, the holder is thoroughly cleaned with water and ethanol and dried for the next experiment. 

\subsection*{Cell for ionic conductivity measurement}
The steel electrode for ionic conductivity measurement is CNC-machined to ensure accurate dimensions. Nylon spacer is purchased from McMaster Carr (SKU\# 94639A694). After measurement, the cell is disassembled and cleaned for reuse.
\section{Calculation of all the pairwise combinations}
\label{si-note:pw-calculation}
In total, there are 171 pairwise combinations of all 19 metals used in this study. The number is calculated as 
\begin{equation}
    C_{19}^2 = \frac{19\times18}{2}= 171
\end{equation}
\section{Calculation of Shannon surprise for agents' experiment design}
\label{si-note:shannon-surprise}
Given a probabilistic model that assigns likelihood $p(y\mid\mathcal{D}, x)$ 
to an observed outcome $y$ conditioned on prior data $\mathcal{D}$, the Shannon surprise (self-information) is defined as
\begin{equation}
    S(y \mid \mathcal{D}, x) 
    = - \log p(y \mid \mathcal{D}, x).
\end{equation}

In this work, we estimate $p(y \mid \mathcal{D}, x)$ using a Gaussian process (GP) regression model (provided by Ax package \cite{olson2025ax}) trained on all previously acquired experimental data prior to each new experiment proposal. Each experiment is encoded as a composition-temperature feature vector. Elemental identities and stoichiometries are represented using one-hot encodings derived from matminer’s \texttt{ElementFraction} featurizer~\cite{ward2018matminer}. The maximum synthesis temperature is discretized into 50 $ ^\circ$C bins and encoded as one-hot vectors. The objective to predict is the measured ionic conductivity. Principal component analysis (PCA) is applied to the combined composition-temperature feature matrix, retaining components that explain 95\% of the variance. The GP model employs a Mat\'ern kernel with $\nu = 5/2$ and a fixed random seed (42) to ensure reproducibility.

\section{Samples in initial human test}
\label{si-note:initial-human-test}
The initial grid test refers to samples tested at the beginning of the campaign or submitted by humans during the campaign. 

Supplementary Table \ref{si-table:initial-test} shows all the groups of compositions explored in the initial human test. We perform a grid search between divalent species (\ce{Mg^{2+}}, \ce{Zn^{2+}}, \ce{Fe^{2+}}, \ce{Mn^{2+}}, \ce{Co^{2+}}, \ce{Ni^{2+}}, \ce{Cr^{2+}}), and trivalent (\ce{V^{3+}}, \ce{Cr^{3+}}, \ce{Y^{3+}}, \ce{In^{3+}}, \ce{Sc^{3+}}) and tetravalent/pentavalent (\ce{Zr^{4+}}, \ce{Nb^{5+}}, \ce{Ta^{5+}}). This search does not include all possible binary combinations, but should include all common combinations of metals reported in the spinel space. Additionally, we also tested three other scenarios with fewer samples to ensure the stability of the system: (1) we randomly generate 14 ternary compounds from all the available metal species to make sure the robot is able to process samples across different chemistries; (2) 8 samples with stoichiometry of metal more than 1, which represents a promising doping strategy reported in the recent literature \cite{baumgartner_highly_2025}; (3) 4 samples that human thinks promising in the Mn(II)-V(III) system.

The synthesis temperature is manually selected based on prior scientific literature and the melting points of the metal chloride precursors, typically set to 400 or 450 $^\circ$C. These samples constitute a human-driven campaign that combines a grid search within regions deemed promising by domain expertise, together with targeted compositions identified as particularly promising based on prior knowledge and intuition.

\section{Training Bayesian optimization model for BO-assisted pattern-finding agent}
\label{si-note: bo-training}
The Bayesian optimization model is trained with Ax \cite{olson2025ax} and BoTorch \cite{balandat2020botorch} packages. A dataset of (composition, purity, and conductivity) is first constructed. If multiple samples with the same composition are present, the one with the highest ionic conductivity is used as a representative sample. The composition is encoded using MatScholar elemental embedding, trained on the material science literature \cite{tshitoyan2019unsupervised}, as implemented in MatMiner \cite{ward2018matminer}, to capture element similarity. Principal component analysis (PCA) is applied to the compositional embedding matrix, retaining components that explain 90\% of the variance, followed by a standardization to normalize each dimension to unit norm. The GP model employs a Mat\'ern kernel with $\nu = 5/2$ and a fixed random seed (42) to ensure reproducibility. The Gaussian process model is trained to predict both the phase content of the spinel phase and ionic conductivity at the same time. A joint purity-conductivity score $S_\text{joint}$ along with its variance is defined as
\begin{equation}
    S_\text{joint}=\log \hat{\sigma} + \lambda_p \cdot \hat{p}\cdot \mathbf{1}_{\hat{p} > 0.9}(\hat{p})
\end{equation}
\begin{equation}
    \text{Var}(S_\text{joint}) \approx \text{Var}(\log \hat{\sigma}) + \lambda_p^2 \cdot \text{Var}(\hat{p})
\end{equation}
where $\hat{\sigma}$ is the predicted ionic conductivity, $\hat{p}$ is the predicted phase content of spinel phase in the sample, and $\lambda_p$ is the weighting parameter to adjust the contribution of phase purity to the score, which is set to 5.0 by default. $\mathbf{1}_{\hat{p} > 0.9}(\hat{p})$ is the indicator function, where samples with a high spinel fraction (\textgreater90wt\%) have a bonus score to the $S_\text{joint}$. The variance of $S_\text{joint}$ is approximated by ignoring the variance of the indicator function ($\mathbf{1}_{\hat{p} > 0.9}(\cdot)$) in the definition, which is enough for capturing the uncertainty in the phase purity.

A large discrete pool of candidate compositions (5,000) is generated by random sampling over a predefined set of metal cations with specified oxidation states: \ce{Mg^{2+}}, \ce{V^{3+}}, \ce{Cr^{2+}}, \ce{Cr^{3+}}, \ce{Mn^{2+}}, \ce{Fe^{2+}}, 
\ce{Co^{2+}}, \ce{Ni^{2+}}, \ce{Cu^{+}},\ce{Cu^{2+}}, \ce{Zn^{2+}}, \ce{Y^{3+}}, \ce{Zr^{4+}}, \ce{Nb^{5+}}, \ce{Hf^{4+}}, \ce{Ta^{5+}}, \ce{In^{3+}}, \ce{Sn^{2+}}, \ce{Bi^{3+}}. In each composition, 4-6 species are randomly selected, with stoichiometry ranging from 0.1 to 1.0 in increments of 0.1. Each composition is required to have $x\in[0, 0.8]$ and $y\in[-0.1, 0.2]$ in the formula \ce{Li_{2-x}M_{1±$y$}Cl4} to be a valid composition.

Once the candidate compositions are generated, the trained model calculates an upper confidence bound (UCB) acquisition function, defined as
\begin{equation}
    a_\text{UCB} = S_\text{joint}+\lambda_\text{UCB} \sqrt{\text{Var}(S_\text{joint})}-\beta \cdot \text{NElem}
\end{equation}
where NElem is the number of elements in each proposed composition as a complexity penalty to the proposed composition, $\lambda_\text{UCB}$ is defined as the weighing parameter to adjust the ratio between exploration and exploitation, and is set to 1.0, and $\beta$ is the weighing parameter for the complexity penalty, which is set to 1.0 in this study.

To avoid redundant proposals and promote chemical diversity, a weighted k-determinantal point process (k-DPP) \cite{grosse2024greedy} is applied to the candidate pool to determine the short list of candidate compositions sent to the BO-assisted pattern-finding agent. A radial basis function (RBF) kernel is used to estimate the distance between compositions, where each composition is encoded as a one-hot vector ($\mathbf{c}$).

\begin{equation}
    k_{ij}=\exp\left(-\frac{\|\mathbf{c}_i-\mathbf{c}_j\|^2}{2l^2}\right)
\end{equation}

where $l$ is the length scale, set to 1 in this study. 

To obtain the overall diversity while maintaining a good acquisition function score $a_\text{UCB}$, we define the quality weight $L_{ij}$ as
\begin{equation}
    L_{ij} = \exp(a_{\text{UCB},i}-\max(a_\text{UCB}))\cdot \exp(a_{\text{UCB},j}-\max(a_\text{UCB}))\cdot k_{ij}
\end{equation}

The exponential operation offset by the max $a_\text{UCB}$ is to ensure $L_{ij}$ is positive. With a subset $S$ of $N$ samples ($N=40$ in this study), it can form a matrix $L_S$. The determinant value of this matrix $\det(L_S)$ can be interpreted as the volume of space this matrix occupies. To maximize the diversity of selection, the objective function to find a subset $S^*$, s.t.,
\begin{equation}
    S^*=\arg \max_{|S|=N}\det(L_S)
\end{equation}

To find such a subset $S^*$, we employ a greedy algorithm, where it starts with the composition that has the highest $a_\text{UCB}$. At each step, it calculates the improvement $\Delta(i|S_{i-1})$ for each candidate composition as
\begin{equation}
    \Delta(i|S_{i-1})=\det(L_{S_{i-1}\cup\{i\}})-\det(L_{S_{i-1}})
\end{equation}

The candidate list is returned once the target number is reached.

\section{Sanity checks for agent-proposed experiment}
\label{si-note:sanity-check}
Before submitting the agent-generated experiments to A-Lab GPSS, a few checks are run before the submission.
\begin{itemize}
    \item \textbf{Valid composition}: The proposed composition must be made with available precursors and be charge-balanced. If not, it is rejected instantly.
    \item \textbf{Rounding heating temperature}: For all agents, the heating temperature is rounded to the nearest 50 $^\circ$C to ensure more samples can be heated together in one heating cycle.
    \item \textbf{Constraining heating temperature}: For the abnormality-detection agent, the agent can propose the synthesis temperature for synthesizing a composition. However, too high a temperature can cause halide evaporation, leading to the purifier in the glovebox being poisoned. To constrain the temperature, we enforce that the proposed temperature should not exceed 200$^\circ$C plus 3/4 of the average melting point of all the precursors ($T_\text{modified Tammann}+200$).
    \item \textbf{Restricted usage of \ce{ScCl3}}: \ce{ScCl3} is a costly and precious precursor, although incorporation of \ce{Sc^{3+}} typically enhances ionic conductivity. Due to the limited inventory of \ce{ScCl3}, its usage is limited to 200~mg per experiment, within a total batch mass of 500~mg per synthesis. Any composition that requires a high amount of \ce{ScCl3} is rejected. Furthermore, to preserve the remaining supply, \ce{Sc^{3+}} is prohibited for the two pattern-finding agents when proposing new compositions.
    \item \textbf{No duplicate experiments}: Duplicate experiments are not permitted. For the abnormality-detection agent, a proposed experiment is rejected if the same composition has previously been tested at the same synthesis temperature. For the two pattern-finding agents, a proposal is rejected if the composition has already been explored, since these agents cannot modify the synthesis temperature.
\end{itemize}

\section{Changes made to the analysis workflow during the campaign}
\label{si-note:changes}
\subsection{XRD phase analysis}
The XRD analysis workflow used slightly different fitting parameters for the first two batches of the campaign. During this initial stage, all phases were allowed to have a peak broadening parameter B\textsubscript{1} in the range of 0 to 0.08, which in some cases led to incomplete phase identification. This issue was identified and corrected after the second batch, resulting in the analysis workflow described in the Methods section. (Only target spinel phase's B\textsubscript{1} is allowed 0-0.08, while others are only 0-0.01)

\subsection{Ionic conductivity analysis}
During the campaign, the ionic conductivity values of 13 samples were updated. 
First, in the initial workflow, samples with very low apparent ionic conductivity, particularly those without a clear semicircle in the Nyquist plot, were assigned a default value of $1 \times 10^{-8}$ S/cm to indicate low conductivity. Upon later inspection, we found that in most of these cases, the semicircle was still weakly present but obscured by noise, making a visual estimate of the impedance more appropriate. This affected samples 4, 8, 11, 18, 26, and 43, as listed in Supplementary Table~\ref{si-tab:changes}. Most of the samples are updated after the second batch in the campaign.
Second, we identified a small number of cases in which the evolutionary fitting algorithm reached an RMSE below 1.0 but nevertheless did not produce a physically reasonable impedance fit. These spectra were manually reviewed, and the corresponding ionic conductivity values were corrected accordingly. This affected samples with index 10, 33, 78, 113, 125, 180, and 251, as shown in Supplementary Table~\ref{si-tab:changes}.

\section{Claim extraction from agents' reasoning processes}
\label{si-note:claim-extraction}
The claim-extraction workflow is implemented with OpenAI's GPT-5-mini, with medium reasoning efforts. A claim is defined as a statement asserting a fact that can be independently verified. Each claim must contain a single verifiable assertion and include sufficient contextual information to enable verification. Extracted claims are classified into two categories. \textbf{Dataset-referenced claims} refer explicitly to experimental results within the A-Lab GPSS dataset, typically specifying a composition (or a narrowly defined composition range) together with quantitative values (e.g., ionic conductivity or phase fraction). Because the dataset is structured and limited to XRD and ionic conductivity measurements, LLMs can accurately identify such claims. \textbf{External-referenced claims} rely on scientific knowledge beyond the internal dataset, which cannot be made directly from the dataset.


\section{Experiments related to Li--Fe--Mn--In--Y--Cl chemical system}
\label{si-note:li-fe-mn-in-y-cl}

\input{auto-gen-si/li_fe_mn_in_y_cl}

\section{Experiments related to Li--Mn--Sc--Cl chemical system}
\label{si-note:li-mn-sc-cl}
A summary of experiments on the Li--Mn--Sc--Cl chemical system is provided.

\input{auto-gen-si/li-mn-sc-cl}

\section{\texorpdfstring{Experiment trajectory for proposing Li\textsubscript{1.8}Mn\textsubscript{1.1}Cl\textsubscript{4}}{Experiment trajectory for proposing Li1.8Mn1.1Cl4}}
\label{si-note:li-mn-cl}
The trajectory of exploration (sorted by sample index) for agents to reach the \ce{Li_{1.8}Mn_{1.1}Cl4} composition is shown below.

\input{auto-gen-si/li1p8mn1p1cl4}

\newpage
\renewcommand{\arraystretch}{1.5}
\begin{longtable}{p{1.5in}|>{\raggedright\arraybackslash}p{5in}}
\caption{All the samples that are included in the initial human test.}
\label{si-table:initial-test}\\
\hline
\textbf{Group} & \textbf{Compositions} \\
\hline
\endfirsthead

\hline
\textbf{Group} & \textbf{Compositions} \\
\hline
\endhead

\hline
\endfoot

Single elements (Divalent) &
\ce{Li2MgCl4}, \ce{Li2ZnCl4}, \ce{Li2FeCl4}, \ce{Li2MnCl4},
\ce{Li2CoCl4}, \ce{Li2NiCl4}, \ce{Li2CrCl4} \\
\hline

Isovalent doping (Divalent) &
\ce{Li2Co_{0.5}Fe_{0.5}Cl4}, \ce{Li2Mg_{0.5}Fe_{0.5}Cl4},
\ce{Li2Mn_{0.5}Fe_{0.5}Cl4}, \ce{Li2Co_{0.5}Zn_{0.5}Cl4},
\ce{Li2Co_{0.5}Mg_{0.5}Cl4}, \ce{Li2Co_{0.5}Mn_{0.5}Cl4},
\ce{Li2Mg_{0.5}Mn_{0.5}Cl4}, \ce{Li2Zn_{0.5}Mg_{0.5}Cl4},
\ce{Li2Zn_{0.5}Mn_{0.5}Cl4}, \ce{Li2Fe_{0.5}Zn_{0.5}Cl4},
\ce{Li2Fe_{0.8}Mn_{0.2}Cl4}, \ce{Li2Fe_{0.8}Ni_{0.2}Cl4},
\ce{Li2Cr_{0.5}Mn_{0.5}Cl4} \\
\hline

Trivalent &
\ce{Li_{1.5}Fe_{0.5}V_{0.5}Cl4}, \ce{Li_{1.5}Mn_{0.5}V_{0.5}Cl4},
\ce{Li_{1.5}Co_{0.5}V_{0.5}Cl4}, \ce{Li_{1.5}Mg_{0.5}V_{0.5}Cl4},
\ce{Li_{1.5}Fe_{0.5}Cr_{0.5}Cl4}, \ce{Li_{1.5}Mn_{0.5}Cr_{0.5}Cl4},
\ce{Li_{1.5}Co_{0.5}Cr_{0.5}Cl4}, \ce{Li_{1.5}Mg_{0.5}Cr_{0.5}Cl4},
\ce{Li_{1.5}Fe_{0.5}Y_{0.5}Cl4}, \ce{Li_{1.5}Mn_{0.5}Y_{0.5}Cl4},
\ce{Li_{1.5}Co_{0.5}Y_{0.5}Cl4}, \ce{Li_{1.5}Mg_{0.5}Y_{0.5}Cl4},
\ce{Li_{1.8}Fe_{0.8}In_{0.2}Cl4},
\ce{Li_{1.8}Co_{0.8}In_{0.2}Cl4}, \ce{Li_{1.8}Mg_{0.8}In_{0.2}Cl4},
\ce{Li_{1.8}Mn_{0.8}In_{0.2}Cl4}, 
\ce{Li_{1.5}Fe_{0.5}Sc_{0.5}Cl4}, \ce{Li_{1.5}Mn_{0.5}Sc_{0.5}Cl4},
\ce{Li_{1.5}Co_{0.5}Sc_{0.5}Cl4}, \ce{Li_{1.5}Mg_{0.5}Sc_{0.5}Cl4} \\
\hline

Tetravalent / Pentavalent &
\ce{Li_{1.6}Mn_{0.8}Zr_{0.2}Cl4}, \ce{Li_{1.6}Co_{0.8}Zr_{0.2}Cl4},
\ce{Li_{1.6}Zn_{0.8}Zr_{0.2}Cl4}, \ce{Li_{1.6}Mg_{0.8}Zr_{0.2}Cl4},
\ce{Li_{1.6}Fe_{0.8}Zr_{0.2}Cl4}, \ce{Li_{1.4}Mn_{0.8}Nb_{0.2}Cl4},
\ce{Li_{1.4}Co_{0.8}Nb_{0.2}Cl4}, \ce{Li_{1.4}Zn_{0.8}Nb_{0.2}Cl4},
\ce{Li_{1.4}Mg_{0.8}Nb_{0.2}Cl4}, \ce{Li_{1.4}Mn_{0.8}Ta_{0.2}Cl4},
\ce{Li_{1.4}Co_{0.8}Ta_{0.2}Cl4}, \ce{Li_{1.4}Zn_{0.8}Ta_{0.2}Cl4},
\ce{Li_{1.4}Mg_{0.8}Ta_{0.2}Cl4}, \ce{Li_{1.4}Fe_{0.8}Nb_{0.2}Cl4},
\ce{Li_{1.4}Fe_{0.8}Ta_{0.2}Cl4} \\
\hline

Randomly generated (involving less common species) &
\ce{Li_{1.4}Cr_{0.6}Fe_{0.2}Ta_{0.2}Cl4}, \ce{Li_{1.6}Cr_{0.4}Mn_{0.2}Y_{0.4}Cl4},
\ce{Li_{1.8}Cr_{0.6}Fe_{0.2}Y_{0.2}Cl4}, \ce{Li_{1.6}Ca_{0.2}Mg_{0.4}Y_{0.4}Cl4},
\ce{Li_{1.6}Co_{0.4}Cr_{0.4}Zn_{0.2}Cl4}, \ce{Li2Co_{0.4}Fe_{0.2}Ni_{0.4}Cl4},
\ce{Li_{1.6}In_{0.2}Mg_{0.6}Sc_{0.2}Cl4}, \ce{Li_{1.4}Mg_{0.2}Ni_{0.2}Y_{0.6}Cl4},
\ce{Li_{1.4}Ca_{0.6}Co_{0.2}Ta_{0.2}Cl4}, \ce{Li_{1.4}Fe_{0.2}Mg_{0.6}Ta_{0.2}Cl4},
\ce{Li_{1.4}Ca_{0.6}Mg_{0.2}Nb_{0.2}Cl4}, \ce{Li2Ca_{0.2}Mg_{0.6}Sn_{0.2}Cl4},
\ce{Li_{1.8}Bi_{0.2}Ca_{0.2}Mg_{0.6}Cl4}, \ce{Li_{1.6}Ni_{0.4}V_{0.4}Zn_{0.2}Cl4} \\
\hline

Metal excess &
\ce{Li_{1.6}Fe_{0.9}V_{0.2}Cl4}, \ce{Li_{1.6}Fe_{0.9}Cr_{0.2}Cl4},
\ce{Li_{1.6}Fe_{0.8}Mn_{0.4}Cl4}, \ce{Li_{1.6}Fe_{0.8}Co_{0.4}Cl4},
\ce{Li_{1.6}Fe_{0.8}Mg_{0.4}Cl4}, \ce{Li_{1.6}Fe_{0.9}Y_{0.2}Cl4},
\ce{Li_{1.6}Fe_{0.9}Sc_{0.2}Cl4}, \ce{Li_{1.4}Cr_{1.2}Cl4} \\
\hline

Mn-V mini-campaign &
\ce{Li_{1.65}Mn_{0.95}V_{0.15}Cl4}, \ce{Li_{1.7}MnV_{0.1}Cl4},
\ce{Li_{1.2}Mn_{0.5}V_{0.6}Cl4}, \ce{Li3VCl6} \\
\hline

\end{longtable}

\begin{table}[htbp]
\centering
\caption{Sources of metal chloride precursors used in this study.}
\label{si-tab:precursors}
\begin{tabular}{lll|lll}
\toprule
Formula & Supplier & SKU \# & Formula & Supplier & SKU \# \\
\midrule
LiCl   & Sigma-Aldrich  & 793620     & MgCl$_2$ & Sigma-Aldrich  & 208337     \\
CaCl$_2$ & Sigma-Aldrich  & c4901      & ScCl$_3$ & Sigma-Aldrich  & 409359     \\
VCl$_3$  & Sigma-Aldrich  & 208272     & CrCl$_2$ & Fisher Scientific & AAL1434206 \\
CrCl$_3$ & Fisher Scientific & 5090114566 & MnCl$_2$ & Sigma-Aldrich  & 244589     \\
FeCl$_2$ & Sigma-Aldrich  & 939935     & NiCl$_2$ & Fisher Scientific & AA1468722  \\
CoCl$_2$ & Fisher Scientific & AA1230318  & CuCl     & Sigma-Aldrich  & 212946     \\
CuCl$_2$ & Fisher Scientific & AC206532500 & ZnCl$_2$ & Fisher Scientific & AAA1628122 \\
YCl$_3$  & Sigma-Aldrich  & 451363     & ZrCl$_4$ & Sigma-Aldrich  & 357405     \\
NbCl$_5$ & Sigma-Aldrich  & 510696     & InCl$_3$ & Sigma-Aldrich  & 203440     \\
SnCl$_2$ & Sigma-Aldrich  & 908924     & HfCl$_4$ & Fisher Scientific & AA4561214  \\
TaCl$_5$ & Sigma-Aldrich  & 510688     & BiCl$_3$ & Sigma-Aldrich  & 224839     \\
\bottomrule
\end{tabular}
\end{table}

\begin{table}[htbp]
\centering
\caption{Samples with updated ionic conductivity values during the campaign. Initial indicates the sample is made during the initial test stage.}
\label{si-tab:changes}
\small
\setlength{\tabcolsep}{4pt}
\renewcommand{\arraystretch}{1.2}
\begin{tabular}{cccccc}
\toprule
\makecell[c]{Sample\\Index} &
\makecell[c]{Composition} &
\makecell[c]{Previous Ionic\\Conductivity\\(S/cm)} &
\makecell[c]{Updated Ionic\\Conductivity\\(S/cm)} &
\makecell[c]{Made at\\Batch\\Number} &
\makecell[c]{Update after\\Batch\\Number} \\
\midrule
4   & \ce{Li2Co_{0.5}Zn_{0.5}Cl4}                  & $4.88 \times 10^{-8}$ & $1.10 \times 10^{-7}$ & Initial & 1  \\
8   & \ce{Li2ZnCl4}                                & $1.00 \times 10^{-8}$ & $3.40 \times 10^{-8}$ & Initial & 1  \\
10  & \ce{Li2Zn_{0.5}Mn_{0.5}Cl4}                  & $1.93 \times 10^{-7}$ & $1.00 \times 10^{-8}$ & Initial & 1  \\
11  & \ce{Li2Fe_{0.5}Zn_{0.5}Cl4}                  & $1.00 \times 10^{-8}$ & $3.25 \times 10^{-8}$ & Initial & 1  \\
18  & \ce{Li_{1.4}Zn_{0.8}Nb_{0.2}Cl4}             & $1.00 \times 10^{-8}$ & $2.94 \times 10^{-8}$ & Initial & 1  \\
26  & \ce{Li_{1.4}Zn_{0.8}Ta_{0.2}Cl4}             & $1.00 \times 10^{-8}$ & $4.69 \times 10^{-9}$ & Initial & 1  \\
33  & \ce{Li_{1.4}Fe_{0.8}Nb_{0.2}Cl4}             & $1.00 \times 10^{-8}$ & $1.17 \times 10^{-5}$ & Initial & 1  \\
43  & \ce{Li_{1.5}Mg_{0.5}Cr_{0.5}Cl4}             & $1.00 \times 10^{-8}$ & $1.39 \times 10^{-9}$ & Initial & 11 \\
78  & \ce{Li_{1.6}Co_{0.8}Zr_{0.2}Cl4}             & $1.10 \times 10^{-6}$ & $1.71 \times 10^{-7}$ & Initial & 1  \\
113 & \ce{Li_{1.8}Fe_{0.8}In_{0.2}Cl4}             & $3.54 \times 10^{-6}$ & $2.72 \times 10^{-5}$ & 1       & 3  \\
125 & \ce{Li_{1.4}Mn_{0.4}V_{0.6}Cl4}              & $5.83 \times 10^{-7}$ & $4.18 \times 10^{-6}$ & 3       & 5  \\
180 & \ce{Li_{1.5}Mn_{0.5}Sc_{0.5}Cl4}             & $2.20 \times 10^{-6}$ & $5.32 \times 10^{-5}$ & 3       & 8  \\
251 & \ce{Li_{1.65}Ni_{0.75}Cu_{0.10}Ta_{0.15}Cl4} & $3.25 \times 10^{-5}$ & $1.43 \times 10^{-7}$ & 12      & 13 \\
\bottomrule
\end{tabular}
\end{table}


\begin{figure}[H]
    \centering
    \includegraphics[width=0.25\linewidth]{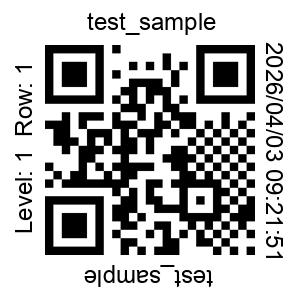}
    \caption{An example tracking label for a sample vial.}
    \label{si-fig:tracking-label}
\end{figure}

\begin{figure}[H]
    \centering
    \includegraphics[width=0.7\linewidth]{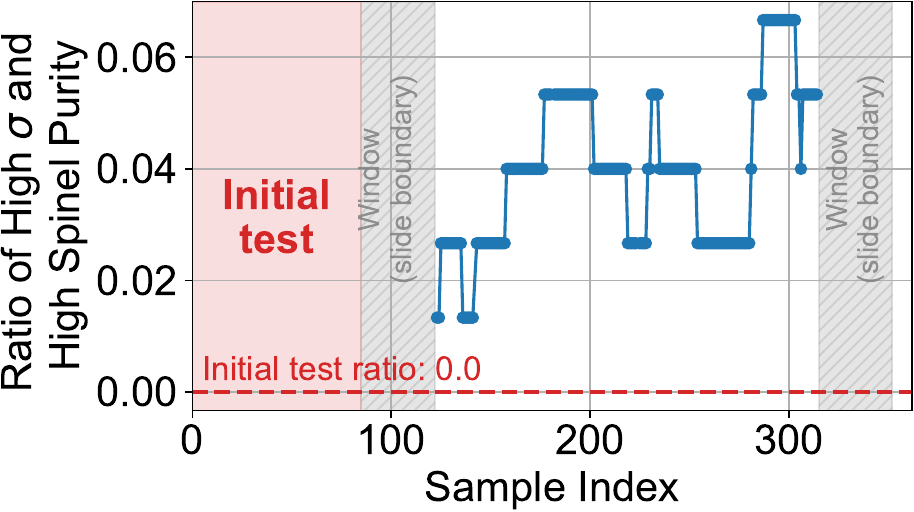}
    \caption{Rolling success rate of high-conductivity, phase-pure spinel samples for AI agents. Samples were considered successful if they showed both high ionic conductivity ($\sigma > 0.05$ mS/cm) and high spinel purity (spinel phase fraction above 80\% and no clear unindexed peaks with relative intensity greater than 0.3 relative to the strongest XRD peak). The $x$-axis shows the sample index, sorted by submission time, and the $y$-axis shows the fraction of successful samples within a rolling window of 75 samples. The red shaded region indicates the initial human-guided test set, whereas the red dashed line marks its success rate ($0.0$). Gray shaded bands indicate the sliding-window boundaries, and the blue line tracks the rolling success rate over time.}
    \label{si-fig:success}
\end{figure}


\newpage

\bibliography{ref}

%% file: auto-gen-si/li_fe_mn_in_y_cl.tex
\subsection*{\ce{Li_{1.5}Mn_{0.5}Y_{0.25}In_{0.25}Cl_{4}} (Index\# 196)}
\begin{itemize}
\item \textbf{Synthesis temperature}: 400~$^\circ$C
\item \textbf{Ionic conductivity}: $1.19 \times 10^{-4}$~S/cm
\item \textbf{XRD results}: spinel (Fd-3m), 100.0 wt\%
\item \textbf{Submitted by}: Pattern

\item \textbf{Justification}: Pattern learned: Mn2+-based inverse spinels with $\sim$0.5 Li vacancies (L$\approx$1.5) and heterovalent M3+ substitution give the highest $\sigma$\_RT among spinel-rich samples (e.g., Li1.5Mn0.5Sc0.5Cl4: 6.43e-05 S/cm; Li1.5Mn0.5Sc0.4V0.1Cl4: 5.2e-05 S/cm), whereas Co/Ni hosts, Zn2+, and any Cr3+ depress conductivity or destabilize the spinel. Y3+ in a Mn host also yields high $\sigma$ (Li1.5Mn0.5Y0.5Cl4: 2.4e-05 S/cm) but can form Li3YCl6; reducing Y content while maintaining the same total 3+ fraction should preserve spinel majority and high conductivity. This design keeps total 3+ = 0.5 (Y0.25+In0.25), giving L = 2 -- ($\Sigma$3+) = 1.5 (y = 0.5). Charge balance: 1.5 (Li+) + 2×0.5 (Mn2+) + 3×0.25 (Y3+) + 3×0.25 (In3+) = 4.0. Mechanism: (i) y$\approx$0.5 creates a percolating 8a–16c–8a Li network with high vacancy concentration; (ii) mixed M3+ (Y3+, In3+) and Mn2+ introduce strong charge/size disorder on 16d that frustrates octahedral ordering (which correlates with low $\sigma$ in Co–Fe systems) and stabilizes Fd-3m; (iii) the higher polarizability of In–Cl compared to purely 3d chlorides softens the lattice and widens migration bottlenecks. Relative to Li1.5Mn0.5Y0.5Cl4, splitting Y with In3+ should suppress Li3YCl6 formation, helping retain \textgreater50\% spinel while targeting $\sigma$\_RT in the high-10$^\wedge$-5 S/cm regime at 400–450 $^\circ$C.
\end{itemize}

\subsection*{\ce{Li_{1.5}Mn_{0.3}Fe_{0.2}Y_{0.25}In_{0.25}Cl_{4}} (Index\# 199)}
\begin{itemize}
\item \textbf{Synthesis temperature}: 400~$^\circ$C
\item \textbf{Ionic conductivity}: $1.40 \times 10^{-4}$~S/cm
\item \textbf{XRD results}: spinel (Fd-3m), 71.3 wt\%; Li6In2.12Cl12 (C12/m1), 28.7 wt\%
\item \textbf{Submitted by}: Pattern

\item \textbf{Justification}: Pattern learned: (a) Mn-based spinels at L$\approx$1.5 and $\Sigma$(3+)$\approx$0.5 achieve the best $\sigma$ (4–6×10$^\wedge$-5 S/cm); (b) mixing multiple 3+ species boosts 16d disorder and conductivity (Sc/V co-doping outperforms single-3+); (c) Fe2+ in spinels can be accommodated without the severe phase segregation seen for Ni/Zn, and Sc-containing Fe samples can be highly conductive. Here $\Sigma$(3+) = 0.5 (Y0.25+In0.25), with total 2+ = 0.5 (Mn0.3+Fe0.2), giving L = 1.5 (y = 0.5). Charge balance: 1.5 (Li+) + 2×(0.3+0.2) (Mn2+/Fe2+) + 3×(0.25+0.25) (Y3+/In3+) = 4.0. Mechanism: (i) y$\approx$0.5 builds a percolating 8a–16c–8a pathway; (ii) four-cation mixing on 16d (Mn, Fe, Y, In) maximizes configurational disorder, suppresses cation ordering (which lowered $\sigma$ in Co–Fe), and broadens the migration bottleneck distribution; (iii) In3+ adds lattice softness vs. purely 3d chlorides, while moderate Fe2+ maintains a 2+ network without the NiCl2/ZnCl2 impurity issues. This composition targets a spinel-majority microstructure with $\sigma$\_RT in the high-10$^\wedge$-5 S/cm range under 400–450 $^\circ$C processing, leveraging the demonstrated Mn+multi-3+ motif while diversifying the 3+ mix to avoid Li3YCl6 dominance.
\end{itemize}

\subsection*{\ce{Li_{1.8}Fe_{0.6}Mn_{0.2}Y_{0.2}Cl_{4}} (Index\# 200)}
\begin{itemize}
\item \textbf{Synthesis temperature}: 400~$^\circ$C
\item \textbf{Ionic conductivity}: $1.83 \times 10^{-5}$~S/cm
\item \textbf{XRD results}: spinel (Fd-3m), 85.7 wt\%; Li3YCl6 (P-3m1), 8.4 wt\%; LiCl (Fm-3m), 5.9 wt\%
\item \textbf{Submitted by}: Pattern

\item \textbf{Justification}: Patterns learned: (i) Moderate Li deficiency (Li$\approx$1.6–1.8) plus aliovalent substitution that preserves Fd-3m favors $\sigma$RT $\sim$1e-5–4e-5 S/cm; (ii) Y3+ at $\sim$0.2 strongly boosts $\sigma$ (Li1.8Fe0.8Y0.2Cl4 → 3.82e-05 S/cm) via lattice softening/expansion; (iii) Mn2+ substitution (0.2–0.4) raises $\sigma$ by enlarging octahedra and promoting Li/M site disorder (e.g., Li1.6Fe0.8Mn0.4Cl4 at 500 $^\circ$C → 2.97e-05 S/cm) while avoiding Ni/Co/Zn-driven ordering. Proposal: Li1.8Fe0.6Mn0.2Y0.2Cl4 is charge-balanced (1.8 + 2×0.6 + 2×0.2 + 3×0.2 = 4) with y=0.2 and M=1.0. Mechanism: Y3+ (octahedral, large radius, high polarizability) expands the Cl framework and lowers Li+ migration barriers along the 16d–48f–16c network; Mn2+ (HS, larger than Fe2+) further widens bottlenecks and frustrates Li/M–vacancy ordering that depresses $\sigma$ in Nb5+/Ni2+ cases. This composition mirrors the high-$\sigma$ features of the Y-only and Mn-rich data while adding compositional disorder to suppress superlattice ordering seen near $\sim$15$^\circ$ 2$\theta$, targeting single-phase Fd-3m at 450–500 $^\circ$C and $\sigma$RT $\geq$ 1e-5 S/cm.
\end{itemize}

\subsection*{\ce{Li_{1.5}Mn_{0.3}Fe_{0.2}Y_{0.25}In_{0.25}Cl_{4}} (Index\# 226)}
\begin{itemize}
\item \textbf{Synthesis temperature}: 450~$^\circ$C
\item \textbf{Ionic conductivity}: $6.18 \times 10^{-5}$~S/cm
\item \textbf{XRD results}: spinel (Fd-3m), 65.2 wt\%; Li3YCl6 (P-3m1), 34.8 wt\%
\item \textbf{Submitted by}: Abnormal

\item \textbf{Justification}: Step-by-step reasoning from the dataset and basic principles: (1) Observation of the outlier: Li1.5Mn0.3Fe0.2Y0.25In0.25Cl4 (400 $^\circ$C) shows spinel (Fd-3m) as the majority phase (0.713) plus a significant Li6In2.12Cl12 (C12/m1) secondary phase (0.287) and delivers the highest $\sigma$\_RT among spinel-majority entries (1.4e-04 S/cm). (2) Relevant comparators: Li1.5Mn0.5Y0.25In0.25Cl4 at 400 $^\circ$C is single-phase spinel and also highly conductive (1.19e-04 S/cm), while Sc-containing spinels at Li$\approx$1.5 yield lower $\sigma$\_RT ($\sim$2.4e-05 to 6.4e-05 S/cm). Thus, high $\sigma$ correlates with Li$\approx$1.5 (vacancy-rich) and Y/In-containing chemistries; the Fe-containing sample is slightly higher than the Y/In-only sample and uniquely includes a sizable Li6In2.12Cl12 secondary phase. (3) General knowledge: Li3MCl6-type chlorides (stoichiometrically related to Li6In2Cl12) are widely reported fast Li+ conductors; in composites, such phases at interparticle contacts can lower pellet-level resistance via fast-grain-boundary pathways and/or space-charge effects. (4) Mechanistic implication: The abnormal $\sigma$ could be boosted by the coexistence of a vacancy/disordered inverse spinel network with a Li3InCl6-type secondary phase that provides low-resistance interfaces; alternatively, the high $\sigma$ could be intrinsic to the quaternary (Mn/Fe/Y/In) disordered inverse spinel at Li$\approx$1.5 independent of the secondary phase. (5) Minimal, decisive test within current lab capabilities: keeping composition fixed and raising the synthesis temperature to 450 $^\circ$C is a realistic way to drive further reaction and potentially reduce or eliminate the Li6In2.12Cl12 fraction (as seen in other systems where impurity fractions change between 400–450 $^\circ$C), without adding new variables. Then, XRD will quantify the secondary-phase fraction and PEIS will measure $\sigma$\_RT. Decision rule: if the Li6In2.12Cl12 fraction substantially decreases (e.g., by $\geq$20 wt\%) and $\sigma$\_RT drops by a comparable fraction ($\geq$25\%) relative to 1.4e-04 S/cm, this supports an interfacial/composite-conduction contribution; if the impurity fraction decreases while $\sigma$\_RT remains $\geq$1.2e-04 S/cm, the high $\sigma$ is likely intrinsic to the disordered, vacancy-rich spinel enabled by multi-cation mixing. Temperature choice: 450 $^\circ$C lies within the 300–700 $^\circ$C range typical for halide spinels, has yielded high-quality spinel products in this dataset, and avoids the $\sigma$ collapse sometimes seen at 500 $^\circ$C in other compositions.
\end{itemize}

\subsection*{\ce{Li_{1.5}Mn_{0.5}Y_{0.25}In_{0.25}Cl_{4}} (Index\# 262)}
\begin{itemize}
\item \textbf{Synthesis temperature}: 450~$^\circ$C
\item \textbf{Ionic conductivity}: $3.98 \times 10^{-5}$~S/cm
\item \textbf{XRD results}: spinel (Fd-3m), 69.8 wt\%; Li3YCl6 (P-3m1), 30.2 wt\%
\item \textbf{Submitted by}: Abnormal

\item \textbf{Justification}: Step-by-step reasoning from the dataset and basic spinel-halide principles: (1) Observation: Li1.5Mn0.5Y0.25In0.25Cl4 made at 400 $^\circ$C is a single inverse spinel (Fd-3m, 1.0 wt fraction; Rwp 5.87\%) with only weak model–data mismatches and shows $\sigma$RT = 1.19e-04 S/cm--well above closely related Mn-based spinel compositions under similar conditions. (2) Controls at similar vacancy level (y$\approx$0.5): Y-only (Li1.5Mn0.5Y0.5Cl4) at 450 $^\circ$C shows significant Li3YCl6 and lower $\sigma$ (2.4e-05 S/cm); Sc-only (Li1.5Mn0.5Sc0.5Cl4) peaks at 450 $^\circ$C (6.43e-05 S/cm) and drops sharply at 500 $^\circ$C (2.2e-06 S/cm); mixed V/Y/Mg at 400 $^\circ$C retains Li3YCl6 and has modest $\sigma$ (2.64e-05 S/cm). Thus, the Y+In equimolar mix at total M3+=0.5 uniquely yields both a clean Fd-3m spinel and the highest $\sigma$. (3) Basic principles: At fixed y=0.5, Li-vacancies provide carriers; multi-cation substitution on 16d generally increases configurational disorder and can flatten the Li+ migration energy landscape. Annealing at higher T can enable cation ordering or phase segregation in spinels, which typically raises migration barriers and/or reduces carrier density. (4) Therefore, the most discriminating, minimal test is to keep composition fixed and change only temperature to 450 $^\circ$C: if the high $\sigma$ at 400 $^\circ$C is tied to kinetically retained cation disorder, a higher-T synthesis (12 h hold) should reduce $\sigma$ and/or introduce secondary phases or stronger model–data mismatches in XRD; if $\sigma$ stays $\approx$1e-4 S/cm with a single Fd-3m phase, the benefit is robust to annealing and mainly compositional (Y/In mixing) rather than a low-T metastable effect. Decision rule: relative to the 400 $^\circ$C baseline, a $\geq$2× $\sigma$ drop with new/stronger secondary phases or increased mismatch features in XRD supports the metastable-disorder hypothesis; comparable $\sigma$ and similar single-phase Fd-3m supports a composition-driven disorder/lattice-softening mechanism.
\end{itemize}

\subsection*{\ce{Li_{1.5}Fe_{0.5}Y_{0.25}In_{0.25}Cl_{4}} (Index\# 282)}
\begin{itemize}
\item \textbf{Synthesis temperature}: 400~$^\circ$C
\item \textbf{Ionic conductivity}: $9.48 \times 10^{-5}$~S/cm
\item \textbf{XRD results}: spinel (Fd-3m), 69.8 wt\%; Li3YCl6 (P-3m1), 18.3 wt\%; LiCl (Fm-3m), 9.9 wt\%; FeCl3 (P-3), 2.0 wt\%
\item \textbf{Submitted by}: Pattern

\item \textbf{Justification}: - Patterns learned from the data:
  - Mixed trivalent Y3+ + In3+ at total 0.5 with Li$\approx$1.5 produced the highest $\sigma$ and a single Fd-3m spinel in a Mn2+ host--evidence that combining two 3+ species can curb Li3YCl6/Li6In2--xCl12 and suppress cation ordering linked to reduced $\sigma$.
  - Heavy 5+ or Cr3+ content generally correlates with ordering/impurities or low $\sigma$; avoiding those while keeping y$\approx$0.5 and mixed 3+ is favorable.
- How this composition follows those patterns:
  - Generalizes the successful Y3+ + In3+ strategy to a different divalent host (Fe2+) to evaluate host-lattice effects on octahedral disorder and Li+ transport. Charge balance holds: 1.5(Li+) + 0.5×2 (Fe2+) + 0.25×3 (Y3+) + 0.25×3 (In3+) = 1.5 + 1.0 + 0.75 + 0.75 = 4.
  - Mechanistic link: size/valence dispersion across 16d (Fe2+ vs. Y3+/In3+) is expected to frustrate long-range order, minimize low-angle superlattice intensity, and preserve a dense vacancy network (y=0.5) that enables 3D migration.
- Why not 100\% sure:
  - The dataset lacks Fe-host Y/In analogs; phase stability vs. formation of Li3YCl6/Li6In2--xCl12 is not known a priori for Fe2+. Fe-based chlorides can have different site preferences and could introduce minor electronic conduction; whether the same high-$\sigma$ mechanism observed in the Mn host carries over is uncertain.
\end{itemize}

\subsection*{\ce{Li_{1.5}Mn_{0.5}Y_{0.1}In_{0.4}Cl_{4}} (Index\# 309)}
\begin{itemize}
\item \textbf{Synthesis temperature}: 450~$^\circ$C
\item \textbf{Ionic conductivity}: $8.65 \times 10^{-6}$~S/cm
\item \textbf{XRD results}: spinel (Fd-3m), 83.8 wt\%; Li6In2.03Cl12 (C12/m1), 16.2 wt\%
\item \textbf{Submitted by}: Abnormal

\item \textbf{Justification}: Step-by-step reasoning from the dataset and basic principles: (1) Observation of the abnormality: Li1.5Mn0.5Y0.25In0.25Cl4 made at 400 $^\circ$C refines as single spinel (Fd-3m, 1.0 wt; Rwp 5.87\%) and shows high $\sigma$RT = 1.19e-04 S/cm. The same nominal composition at 450 $^\circ$C shows phase segregation to Li3YCl6 (spinel 0.6976, Li3YCl6 0.3024) and $\sigma$RT drops to 3.98e-05 S/cm. (2) Corroborating trend with Y: Li1.5Mn0.5Y0.5Cl4 (450 $^\circ$C) also shows significant Li3YCl6 (0.3899) and has low $\sigma$RT = 2.4e-05 S/cm. In contrast, a Y-free analogue Li1.5Mn0.5Sc0.5Cl4 at 450 $^\circ$C is predominantly spinel with no Li3MCl6 phase reported and displays higher $\sigma$RT (6.43e-05 S/cm) than its 400 $^\circ$C counterpart, indicating that the conductivity drop upon going to 450 $^\circ$C is not universal but correlates with Y-containing systems where Li3YCl6 appears. (3) Mechanistic link (basic defect chemistry): In inverse spinels, heterovalent M3+ substitution introduces Li vacancies for charge neutrality; high Li-vacancy content and cation-site disorder generally enhance Li+ transport. If Y3+ partitions out as Li3YCl6 at higher T, the residual spinel loses some heterovalent substitution, reducing Li-vacancy concentration and/or cation disorder, which would increase migration barriers and lower $\sigma$RT. (4) Testable prediction: Reducing the Y fraction while keeping the total trivalent content $\Sigma$(M3+) = 0.5 should suppress Li3YCl6 formation at 450 $^\circ$C and preserve a vacancy-rich disordered spinel, maintaining high $\sigma$RT. (5) Chosen composition and temperature: Li1.5Mn0.5Y0.1In0.4Cl4 keeps $\Sigma$(M3+) = 0.5 (Y0.1+In0.4) to retain $\sim$0.5 Li vacancies while reducing the chemical driving force to form Li3YCl6. Selecting 450 $^\circ$C directly tests whether the conductivity drop at 450 $^\circ$C is tied to Y-driven Li3YCl6 formation rather than an unavoidable temperature-induced ordering. (6) Decision rule using existing lab tools: XRD--if Li3YCl6 peaks are absent and the pattern refines predominantly as spinel (Fd-3m), and $\sigma$RT $\geq$ 1.0e-04 S/cm, the hypothesis is supported. If Li3YCl6 still forms or $\sigma$RT remains $\leq$ $\sim$6e-05 S/cm despite no Li3YCl6, then the primary cause is likely temperature-driven ordering/disorder changes rather than Y partitioning. The design is minimal (one sample), orthogonal (isolates Y content at fixed $\Sigma$(M3+) and fixed temperature), focuses on spinel stability, and uses temperatures proven viable for halide spinels.
\end{itemize}

\subsection*{\ce{Li_{1.5}Fe_{0.5}Y_{0.2}In_{0.3}Cl_{4}} (Index\# 321)}
\begin{itemize}
\item \textbf{Synthesis temperature}: 500~$^\circ$C
\item \textbf{Ionic conductivity}: $7.16 \times 10^{-5}$~S/cm
\item \textbf{XRD results}: spinel (Fd-3m), 76.6 wt\%; Li3YCl6 (P-3m1), 21.0 wt\%; FeCl2 (R-3m), 2.4 wt\%
\item \textbf{Submitted by}: Abnormal

\item \textbf{Justification}: Step-by-step reasoning from the dataset and basic principles: (1) The abnormal sample Li1.5Fe0.5Y0.25In0.25Cl4 (synthesized at 400 $^\circ$C) is multiphase by XRD with spinel (Fd-3m) 0.6975, Li3YCl6 (P-3m1) 0.1834, LiCl (Fm-3m) 0.0989, and FeCl3 (P-3) 0.0203, and shows the highest $\sigma$RT = 9.48e-05 S/cm. (2) A related composition that also contains Li3YCl6, Li1.75Fe0.75Y0.25Cl4 (500 $^\circ$C; spinel 0.8135; Li3YCl6 0.1865), has $\sigma$RT = 4.63e-05 S/cm--elevated versus single-phase spinel samples. (3) Nearly single-phase spinels with comparable Y but no Li3YCl6 (e.g., Li1.78Fe0.78Y0.22Cl4 at 500 $^\circ$C: 97.17 wt\% spinel, $\sigma$RT = 4.23e-05 S/cm; Li1.8Fe0.8Y0.2Cl4 at 500 $^\circ$C: 100 wt\% spinel, $\sigma$RT = 3.82e-05 S/cm) exhibit lower $\sigma$RT than the Li3YCl6-containing mixtures. (4) A Y/Mn co-doped product with only 8.4 wt\% Li3YCl6 (Li1.8Fe0.6Mn0.2Y0.2Cl4 at 400 $^\circ$C) gives $\sigma$RT = 1.83e-05 S/cm. (5) From general, widely accepted knowledge: Li3YCl6 is a recognized fast Li+ conductor in halide systems, whereas LiCl at room temperature is a poor Li+ conductor; thus LiCl is unlikely to account for the high $\sigma$RT. (6) Therefore, the most parsimonious, testable interpretation is that the elevated $\sigma$RT in Li1.5Fe0.5Y0.25In0.25Cl4 is dominated by parallel Li+ transport through the Li3YCl6 fraction rather than being intrinsic to the spinel matrix. (7) A decisive test is to suppress Li3YCl6 while keeping the Li-vacancy level (y) fixed and the total M3+ content constant; reducing Y while increasing In keeps y = 0.5 constant but moves Y below the apparent Li3YCl6-formation window suggested by this dataset (Y = 0.25 at 500 $^\circ$C produced Li3YCl6; Y = 0.22 at 500 $^\circ$C did not). The prior Y+In sample did not show any In-based Li–In chloride by XRD, so lowering Y (while increasing In) specifically targets suppression of Li3YCl6. (8) Chosen composition Li1.5Fe0.5Y0.2In0.3Cl4 maintains charge balance in the Li2--yM1Cl4 framework with y = 0.5 and M = Fe0.5Y0.2In0.3 = 1.0. (9) Temperature choice: 500 $^\circ$C is justified because, in this dataset, 500 $^\circ$C yielded nearly/single-phase spinels at similar Y levels (0.20–0.22) and suppressed Li3YCl6 compared with lower-temperature syntheses. Decision rule: If XRD shows no Li3YCl6 and $\sigma$RT falls to the spinel-only range observed in this dataset ($\sim$few×10$^\wedge$-5 S/cm), this supports the hypothesis that Li3YCl6 dominated the abnormal high $\sigma$. If Li3YCl6 is absent yet $\sigma$RT remains near $\sim$1×10$^\wedge$-4 S/cm, the high conductivity is likely intrinsic to the highly Li-deficient, Y/In co-doped inverse spinel. If Li3YCl6 persists, the effective Y solubility in the presence of In is lower than inferred, and further Y reduction would be warranted.
\end{itemize}

\subsection*{\ce{Li_{1.5}Mn_{0.25}Fe_{0.25}Y_{0.25}In_{0.25}Cl_{4}} (Index\# 324)}
\begin{itemize}
\item \textbf{Synthesis temperature}: 400~$^\circ$C
\item \textbf{Ionic conductivity}: $1.36 \times 10^{-4}$~S/cm
\item \textbf{XRD results}: spinel (Fd-3m), 81.3 wt\%; Li3YCl6 (P-3m1), 18.7 wt\%
\item \textbf{Submitted by}: BO-assisted

\end{itemize}

\subsection*{\ce{Li_{1.5}Fe_{0.5}Y_{0.25}In_{0.25}Cl_{4}} (Index\# 345)}
\begin{itemize}
\item \textbf{Synthesis temperature}: 500~$^\circ$C
\item \textbf{Ionic conductivity}: $9.31 \times 10^{-5}$~S/cm
\item \textbf{XRD results}: spinel (Fd-3m), 77.8 wt\%; Li3YCl6 (P-3m1), 12.0 wt\%; LiCl (Fm-3m), 10.2 wt\%
\item \textbf{Submitted by}: Abnormal

\item \textbf{Justification}: Step-by-step reasoning from the data and basic principles: (1) What is observed: The abnormal sample Li1.5Fe0.5Y0.25In0.25Cl4 synthesized at 400 $^\circ$C is multiphase by XRD: spinel (Fd-3m) 0.6975, Li3YCl6 (P-3m1) 0.1834, LiCl (Fm-3m) 0.0989, FeCl3 (P-3) 0.0203; it shows an extra low-angle peak (15.82$^\circ$) and several missing peaks relative to the spinel model, and $\sigma$RT = 9.48e-05 S/cm (highest among the listed samples). (2) Comparison set: Spinel-only or spinel-dominant Y-containing references at 500 $^\circ$C show lower $\sigma$RT: Li1.8Fe0.8Y0.2Cl4 (Fd-3m 1.0) 3.82e-05; Li1.78Fe0.78Y0.22Cl4 (Fd-3m 0.9717 + LiCl 0.0283) 4.23e-05; Li1.7Fe0.85Y0.2Cl4 (Fd-3m 1.0) 7.64e-05; while Li1.75Fe0.75Y0.25Cl4 containing Li3YCl6 (0.1865) reaches 4.63e-05. Across these, the presence of Li3YCl6 coincides with increased $\sigma$ relative to single-phase spinel baselines. (3) Widely accepted knowledge: Li3YCl6 is a fast Li-ion conductor, so even a moderate fraction can raise bulk pellet $\sigma$ via parallel Li+ transport pathways. (4) Interpretation constrained by XRD: The abnormal sample contains $\sim$18 wt\% Li3YCl6; this correlates with the highest $\sigma$RT observed here. The extra low-angle feature signals deviations from the ideal spinel model but does not by itself quantify transport. (5) Minimal, decisive test: Change only synthesis temperature for the same nominal composition. If the high $\sigma$ originates from Li3YCl6 retained at 400 $^\circ$C (incomplete reaction/limited solubility at low T), raising T to 500 $^\circ$C (used successfully for peers) should reduce Li3YCl6 and other by-products, shifting the product toward a more single-phase spinel. Decision rule with available tools: After synthesizing at 500 $^\circ$C, (i) if XRD shows Li3YCl6 strongly reduced/absent and $\sigma$RT drops toward the spinel-only band observed for peers ($\approx$3.8e-05 to 7.6e-05 S/cm), it supports the hypothesis that Li3YCl6 dominated the abnormal $\sigma$; (ii) if Li3YCl6 persists and $\sigma$ remains near $\sim$1e-4 S/cm, the abnormality is still explained by the secondary phase; (iii) if Li3YCl6 is suppressed yet $\sigma$ remains $\sim$1e-4 S/cm, then the spinel matrix with Y+In may intrinsically support higher $\sigma$, falsifying the hypothesis. Temperature choice (500 $^\circ$C) is within the demonstrated range for halide spinels here and maximizes reaction completion while aligning with prior single-phase spinel syntheses.
\end{itemize}

\subsection*{\ce{Li_{1.5}Mn_{0.3}Fe_{0.2}Y_{0.35}In_{0.15}Cl_{4}} (Index\# 350)}
\begin{itemize}
\item \textbf{Synthesis temperature}: 400~$^\circ$C
\item \textbf{Ionic conductivity}: $1.11 \times 10^{-4}$~S/cm
\item \textbf{XRD results}: spinel (Fd-3m), 61.7 wt\%; Li3YCl6 (P-3m1), 33.2 wt\%; FeCl2 (Pa-3), 5.1 wt\%
\item \textbf{Submitted by}: Abnormal

\item \textbf{Justification}: Step-by-step reasoning from data and basic principles:
1) What is abnormal: Li1.5Mn0.3Fe0.2Y0.25In0.25Cl4 at 400 $^\circ$C is a two-phase mixture (spinel (Fd-3m) 0.713 + Li6In2.12Cl12 (C12/m1) 0.287; Rwp 5.91\%) yet shows $\sigma$RT = 1.4e-04 S/cm. At 450 $^\circ$C the same nominal composition yields spinel 0.6516 + Li3YCl6 0.3484 and a lower $\sigma$RT = 6.18e-05 S/cm. In contrast, Y-only analogues without In (Li1.5Mn0.5Y0.5Cl4, Li1.8Mn0.8Y0.2Cl4) show lower $\sigma$RT ($\sim$10$^\wedge$-5 S/cm). A related Y/In composition without Fe (Li1.5Mn0.5Y0.25In0.25Cl4) is single-phase spinel at 400 $^\circ$C and also highly conductive (1.19e-04 S/cm).
2) Basic knowledge applied: (i) Trivalent substitution (Y3+, In3+) in an inverse spinel Li2--yM1Cl4 host introduces Li vacancies (lower L) that generally enhance Li+ mobility by enlarging the accessible network. (ii) Li3MCl6-type phases, particularly In-based, are widely reported as good Li-ion conductors; a minor, well-dispersed fraction can lower pellet-scale resistance if it occupies intergranular regions. (iii) Changing synthesis temperature can change secondary-phase identity and cation distribution.
3) Linking observations to a mechanism: The 400 $^\circ$C Fe-containing sample pairs a vacancy-rich/disordered spinel with an In-rich Li–In chloride secondary phase and attains the highest $\sigma$RT (1.4e-04 S/cm). At 450 $^\circ$C, the secondary phase identity shifts to Y-rich Li3YCl6 and $\sigma$RT drops. Y-only compositions (no In) at comparable temperatures show systematically lower $\sigma$RT. This pattern is consistent with an added contribution from an In-containing Li–halide secondary phase at 400 $^\circ$C boosting macroscopic $\sigma$ beyond the spinel-only baseline.
4) Testable lever to isolate the effect: At fixed total M3+ = 0.5 (to keep the Li-vacancy concentration constant at L $\approx$ 1.5), decrease In and increase Y to suppress formation of Li–In chloride secondary phases while keeping Mn/Fe2+ and total 3+ content unchanged. If the composite (secondary-phase) contribution is important, $\sigma$RT should drop when Li–In chloride is suppressed; if $\sigma$ remains high, the high $\sigma$ is intrinsic to the disordered, vacancy-rich spinel.
5) Chosen composition and temperature: Li1.5Mn0.3Fe0.2Y0.35In0.15Cl4 preserves charge balance and the same total trivalent content (Y+In=0.5) as the abnormal sample but reduces In (from 0.25 to 0.15). 400 $^\circ$C is selected because it produced the high-$\sigma$ state in the abnormal sample and single-phase spinel in a closely related composition (Li1.5Mn0.5Y0.25In0.25Cl4). This minimizes confounding by temperature-driven ordering/segregation.
Decision rule (what we learn from one run):
- If XRD shows spinel-only (or no detectable Li–In chloride peaks) and $\sigma$RT $\leq$ 8×10$^\wedge$-5 S/cm, the high $\sigma$ at 400 $^\circ$C is likely composite-enhanced by Li–In chloride in the abnormal sample.
- If XRD shows spinel-only and $\sigma$RT $\geq$ 1.1×10$^\wedge$-4 S/cm, the high $\sigma$ is intrinsic to the vacancy-rich, disordered spinel (secondary phase not required).
- If Li–In chloride peaks persist, the attempted suppression failed; result is inconclusive and suggests In segregation is robust to small compositional changes, guiding a next step of further reducing In or temperature.
All statements are derived from the provided XRD/$\sigma$ data and general, widely accepted knowledge about vacancy effects and Li3MCl6-type halide conductors.
\end{itemize}

%% file: auto-gen-si/li-mn-sc-cl.tex
\subsection*{\ce{Li_{1.5}Mn_{0.5}Sc_{0.5}Cl_{4}} (Index\# 82)}
\begin{itemize}
\item \textbf{Synthesis temperature}: 400~$^\circ$C
\item \textbf{Ionic conductivity}: $4.63 \times 10^{-5}$~S/cm
\item \textbf{XRD results}: spinel (Fd-3m), 92.8 wt\%; MnCl2 (R-3m), 7.2 wt\%
\item \textbf{Submitted by}: Initial test

\end{itemize}

\subsection*{\ce{Li_{1.6}Mn_{0.6}Sc_{0.4}Cl_{4}} (Index\# 96)}
\begin{itemize}
\item \textbf{Synthesis temperature}: 400~$^\circ$C
\item \textbf{Ionic conductivity}: $1.57 \times 10^{-5}$~S/cm
\item \textbf{XRD results}: spinel (Fd-3m), 100.0 wt\%
\item \textbf{Submitted by}: Abnormal

\item \textbf{Justification}: Orthogonal composition test to the abnormal x=0.5 sample. For aliovalent Sc3+ substitution on Mn2+, charge neutrality gives y=x; thus Li1.6Mn0.6Sc0.4Cl4 (x=y=0.4) retains a very high Li-vacancy fraction while reducing Sc content versus x=0.5. If the high conductivity in Li1.5Mn0.5Sc0.5Cl4 is intrinsic to vacancy-rich, disorder-frustrated inverse spinel, Li1.6Mn0.6Sc0.4Cl4 synthesized under the same 400 $^\circ$C protocol should remain single-phase Fd-3m (or near-single-phase) and exhibit similarly elevated $\sigma$\_RT ($\geq$2e-5 S/cm). Conversely, if the abnormal $\sigma$ arises from microstructural percolation between two closely related spinel domains specific to x$\approx$0.5, then at x=0.4 we expect a single spinel lattice and a drop in $\sigma$ to the 1e-6–1e-5 S/cm range. Using 400 $^\circ$C maintains consistency with prior syntheses, minimizes halide volatilization, and preserves any frozen-in cation disorder that is hypothesized to enable high mobility.
\end{itemize}

\subsection*{\ce{Li_{1.5}Mn_{0.5}Sc_{0.5}Cl_{4}} (Index\# 100)}
\begin{itemize}
\item \textbf{Synthesis temperature}: 450~$^\circ$C
\item \textbf{Ionic conductivity}: $6.43 \times 10^{-5}$~S/cm
\item \textbf{XRD results}: spinel (Fd-3m), 97.5 wt\%; ScCl (R-3m), 2.5 wt\%
\item \textbf{Submitted by}: Abnormal

\item \textbf{Justification}: Thermal-history probe of cation (dis)order and microstructural domain effects at the same composition that showed abnormal $\sigma$. A modest increase to 450 $^\circ$C (same ramp/hold, natural cool) should enhance cation mobility enough to evolve any latent Li/M ordering or homogenize the two spinel-like domains observed at 400 $^\circ$C, without excessive halide loss. Decision rule: (a) If new low-angle superlattice intensity (e.g., near $\sim$15$^\circ$ 2$\theta$) appears and $\sigma$\_RT decreases relative to the 400 $^\circ$C sample, this supports the hypothesis that suppressed Li/M ordering and high Li-vacancy content drive the high conductivity; (b) If the two-spinel domain signature collapses to a single spinel lattice yet $\sigma$ remains high or increases, the result argues that microstructural percolation is not required and that intrinsic disorder/vacancy effects dominate; (c) If $\sigma$ drops while XRD indicates increased phase separation or secondary phases, then the abnormal $\sigma$ at 400 $^\circ$C likely benefited from a specific microstructural percolation state that is disrupted by the higher-temperature anneal.
\end{itemize}

\subsection*{\ce{Li_{1.45}Mn_{0.45}Sc_{0.55}Cl_{4}} (Index\# 168)}
\begin{itemize}
\item \textbf{Synthesis temperature}: 450~$^\circ$C
\item \textbf{Ionic conductivity}: $4.17 \times 10^{-5}$~S/cm
\item \textbf{XRD results}: spinel (Fd-3m), 90.0 wt\%; MnCl2 (R-3m), 10.0 wt\%
\item \textbf{Submitted by}: Abnormal

\item \textbf{Justification}: Objective: push $\sigma$\_RT $\geq$ 1e-4 S/cm by increasing the Li-vacancy concentration (y = $\alpha$ for Mn2+/Sc3+), while retaining a disordered Fd-3m inverse spinel. Rationale from data: $\sigma$ rises markedly with Sc3+ (Li1.5Mn0.5Sc0.5Cl4: 6.43e-05 S/cm at 450 $^\circ$C) and falls when Sc is reduced to 0.4 (Li1.6Mn0.6Sc0.4Cl4: 1.57e-05 S/cm), indicating a vacancy/disorder optimum near $\alpha$ $\approx$ 0.5. Hypothesis test: raising Sc fraction slightly to $\alpha$ = 0.55 (y = 0.55) should enhance vacancy-enabled 3D percolation and frustrate Li/M octahedral ordering further, potentially lowering migration barriers. Temperature choice: 450 $^\circ$C previously maintained a disordered spinel without superlattice peaks for $\alpha$ = 0.5 and improved $\sigma$ vs 400 $^\circ$C; using 450 $^\circ$C balances reaction completion and impurity control (minimizing Li3ScCl6) for higher-Sc compositions. Decision rule: if spinel fraction $\geq$ 0.5 and $\sigma$\_RT $\geq$ 1e-4 S/cm, confirm the vacancy/disorder optimum shifts above $\alpha$ = 0.5; if $\sigma$ decreases or Li3ScCl6 dominates, revert focus to $\alpha$ $\approx$ 0.5 as the practical optimum.
\end{itemize}

\subsection*{\ce{Li_{1.5}Mn_{0.5}Sc_{0.5}Cl_{4}} (Index\# 180)}
\begin{itemize}
\item \textbf{Synthesis temperature}: 500~$^\circ$C
\item \textbf{Ionic conductivity}: $5.32 \times 10^{-5}$~S/cm
\item \textbf{XRD results}: spinel (Fd-3m), 100.0 wt\%
\item \textbf{Submitted by}: Abnormal

\item \textbf{Justification}: Test the thermal-disorder hypothesis at a single decisive higher temperature. Prior data show $\sigma$ rises from 4.63e-05 S/cm (400 $^\circ$C, minor MnCl2) to 6.43e-05 S/cm (450 $^\circ$C, clean Fd-3m). A 475 $^\circ$C synthesis will determine if further cation/Li sublattice disorder can be frozen-in during natural cooling, yielding either (a) higher $\sigma$ and no low-angle superlattice features ($\sim$15$^\circ$ 2$\theta$), confirming disorder-driven improvement, or (b) onset of ordering/secondary phases (e.g., LiCl/MnCl2) and reduced $\sigma$, indicating an optimum near 450 $^\circ$C. This single, orthogonal thermal step isolates the variable without changing composition and directly links phase purity/ordering to transport.
\end{itemize}

\subsection*{\ce{Li_{1.6}Mn_{0.6}Sc_{0.4}Cl_{4}} (Index\# 218)}
\begin{itemize}
\item \textbf{Synthesis temperature}: 500~$^\circ$C
\item \textbf{Ionic conductivity}: $3.45 \times 10^{-5}$~S/cm
\item \textbf{XRD results}: spinel (Fd-3m), 91.3 wt\%; LiCl (Fm-3m), 8.7 wt\%
\item \textbf{Submitted by}: Abnormal

\item \textbf{Justification}: Step-by-step from the data: (1) Li1.5Mn0.5Sc0.5Cl4 made at 450 $^\circ$C is nearly single-phase spinel (Fd-3m) and shows high $\sigma$RT=6.43e-05 S/cm with no missing/extra peaks, while at 500 $^\circ$C it still indexes as spinel but exhibits extra\_peaks in the Rietveld output at low 2$\theta$ (calculated peaks that are not observed), and $\sigma$RT drops drastically to 2.2e-06 S/cm. (2) The low-angle reflections in spinel are especially sensitive to cation occupancy/distribution because the scattering contrast between Li and transition/main-group cations strongly affects their intensities; suppression of these reflections relative to an Fd-3m model indicates a change in cation distribution or symmetry relative to the disordered model (general crystallography principle). (3) Order/disorder transitions in spinels are commonly promoted near commensurate cation ratios (e.g., 1:1 mixtures on the octahedral sublattice) and by higher-temperature annealing followed by natural cooling, which can allow ordering to develop; ordering typically reduces Li-sublattice disorder/vacancy percolation and increases migration barriers (basic spinel defect chemistry/transport knowledge). (4) Within the same chemical family, moving away from the 1:1 Mn:Sc ratio (e.g., Li1.6Mn0.6Sc0.4Cl4) still forms the spinel at 400 $^\circ$C with $\sigma$RT=1.57e-05 S/cm (single-phase spinel by XRD), indicating that the framework remains accessible and moderately conducting off the 1:1 point. (5) Therefore, testing Li1.6Mn0.6Sc0.4Cl4 at 500 $^\circ$C isolates whether the severe $\sigma$ drop at 500 $^\circ$C is tied to the 1:1-driven cation ordering seen in Li1.5Mn0.5Sc0.5Cl4 or is a general high-T effect. Decision rule: If Li1.6Mn0.6Sc0.4Cl4 at 500 $^\circ$C shows $\sigma$RT comparable to its 400 $^\circ$C value (within about a factor of 2, i.e., $\geq$8e-06 S/cm) and no suppression of low-angle Fd-3m intensities (few/no low-angle extra\_peaks), this supports the 1:1-specific ordering hypothesis for the abnormal drop. If $\sigma$RT also collapses (e.g., $\leq$3e-06 S/cm) with similar low-angle extra\_peaks, the effect is generic temperature-driven ordering/defect annealing across the series. Temperature choice: 500 $^\circ$C replicates the condition that triggered the abnormality while remaining within the typical 300–700 $^\circ$C window for chloride spinels, giving a clean, orthogonal test without changing the synthesis protocol.
\end{itemize}

\subsection*{\ce{Li_{1.4}Mn_{0.4}Sc_{0.6}Cl_{4}} (Index\# 266)}
\begin{itemize}
\item \textbf{Synthesis temperature}: 500~$^\circ$C
\item \textbf{Ionic conductivity}: $1.28 \times 10^{-4}$~S/cm
\item \textbf{XRD results}: spinel (Fd-3m), 100.0 wt\%
\item \textbf{Submitted by}: Abnormal

\item \textbf{Justification}: Step-by-step reasoning from data and basic principles: (1) Observation at fixed stoichiometry: Li1.5Mn0.5Sc0.5Cl4 shows a \textgreater20× $\sigma$RT drop at 500 $^\circ$C (2.2e-06 S/cm) versus 450 $^\circ$C (6.43e-05 S/cm). Its XRD at 500 $^\circ$C refines as a single spinel (Fd-3m) but the calculated model contains several extra low-angle reflections (e.g., $\sim$15.9–16.7$^\circ$ 2$\theta$) not present in the data, indicating the simple Fd-3m intensity model does not describe the real scattering. (2) Control at similar conditions but off 1:1 ratio: Li1.6Mn0.6Sc0.4Cl4 made at 500 $^\circ$C retains relatively high $\sigma$RT (3.45e-05 S/cm) and shows no low-angle extra\_peaks in the calculated model; only a single high-angle missing peak is reported. (3) Basic spinel knowledge: Inverse spinels commonly exhibit cation ordering on the octahedral sublattice at special cation ratios and elevated temperatures; such ordering changes structure factors and can suppress intensities expected for the disordered Fd-3m model without necessarily producing easily detectable superlattice peaks. Ordering/disorder also impacts Li+ pathway connectivity in the tetrahedral–octahedral network; greater cation disorder and appropriate Li-vacancy populations generally favor higher $\sigma$. (4) Mechanistic link consistent with data: The severe $\sigma$ collapse and model–data intensity mismatch appear only for the 1:1 Mn:Sc sample at 500 $^\circ$C, while an off-stoichiometric sample at the same T maintains high $\sigma$ and a cleaner Fd-3m match. This is consistent with a 1:1-specific octahedral cation ordering (or symmetry-lowering/occupancy pattern) at 500 $^\circ$C that is frustrated when the Mn:Sc ratio deviates from 1:1. (5) Minimal, decisive test: Synthesize the symmetric counterpoint Li1.4Mn0.4Sc0.6Cl4 at 500 $^\circ$C. Charge balance is maintained (1.4 Li+ + 0.4×2+ + 0.6×3+ = +4), preserving the vacancy level comparable to Li1.6Mn0.6Sc0.4Cl4 but with the opposite Mn/Sc skew. Decision rule using available tools: If XRD avoids the low-angle extra\_peaks mismatch seen for the 1:1 sample and $\sigma$RT $\geq$ $\sim$2e-05 S/cm (comparable to Li1.6Mn0.6Sc0.4Cl4 at 500 $^\circ$C), that supports a 1:1-specific ordering origin of the abnormal $\sigma$ drop. If $\sigma$ remains low and the model–data mismatch mirrors the 1:1 case, the abnormality likely arises from another factor (e.g., subtle stoichiometry shifts/defects), prompting a different control in subsequent work. (6) Temperature choice: 500 $^\circ$C is intentionally used because the abnormal behavior manifests at this temperature; halide spinels are typically synthesized between 300–700 $^\circ$C, so 500 $^\circ$C is appropriate for solid-state reaction completion and directly probes the hypothesized ordering regime.
\end{itemize}

\subsection*{\ce{Li_{1.45}Mn_{0.45}Sc_{0.55}Cl_{4}} (Index\# 318)}
\begin{itemize}
\item \textbf{Synthesis temperature}: 500~$^\circ$C
\item \textbf{Ionic conductivity}: $1.44 \times 10^{-4}$~S/cm
\item \textbf{XRD results}: spinel (Fd-3m), 100.0 wt\%
\item \textbf{Submitted by}: Abnormal

\item \textbf{Justification}: Step-by-step reasoning from the dataset and basic principles: (1) What is abnormal: Li1.4Mn0.4Sc0.6Cl4 synthesized at 500 $^\circ$C refines as single spinel (Fd-3m, 1.0 wt; Rwp 9.07\%) with a few model–data mismatches and delivers $\sigma$RT = 1.28e-04 S/cm--well above neighboring compositions and temperatures that mostly fall in the mid-10$^\wedge$-5 S/cm range. (2) Composition–temperature patterns in the data at 500 $^\circ$C: • Li1.5Mn0.5Sc0.5Cl4 (Mn:Sc = 1:1) shows a severe conductivity drop to 2.2e-06 S/cm despite single-phase spinel, with several extra peaks in the calculated model; the companion experiment note for Li1.6 at 500 $^\circ$C interprets this as consistent with a 1:1-specific ordering effect at high T. • Li1.6Mn0.6Sc0.4Cl4 (off 1:1) at 500 $^\circ$C gives 3.45e-05 S/cm (spinel 0.913 + LiCl 0.087). • Li1.4Mn0.4Sc0.6Cl4 (off 1:1) at 500 $^\circ$C gives the highest $\sigma$RT = 1.28e-04 S/cm (single spinel) with a few unindexed and suppressed-calculated peaks, indicating the reference model is imperfect but without assigned second phases. (3) Vacancy count from charge neutrality using typical chloride valences (Mn2+, Sc3+): Li content equals the positive-charge deficit to balance Cl4-. Thus Li1.6, Li1.5, and Li1.4 correspond to vacancy levels y $\approx$ 0.4, 0.5, and 0.6 in Li2–y. More Sc3+ (higher Sc fraction) lowers Li content and increases Li vacancies. (4) Basic solid-state ionics: In spinel-type halides, higher Li-vacancy concentration generally facilitates Li+ hopping by providing more accessible empty sites in the connected tetrahedral–octahedral network; cation ordering on octahedral sites can impede percolation by locally modifying Li-site energies and bottlenecks (widely accepted behavior in spinel frameworks). (5) Linking data to mechanism: At 500 $^\circ$C, the 1:1 sample shows an anomalously low $\sigma$RT, consistent with a composition-specific ordering transition noted in the dataset. Deviating from 1:1 (Li1.6 and Li1.4) avoids this severe drop. Among off-1:1 compositions, $\sigma$RT increases as Li vacancies increase (Li1.6 → Li1.4), with Li1.4 giving the highest $\sigma$RT. The minor XRD misfits for Li1.4 indicate the actual cation distribution differs somewhat from the ideal Fd-3m reference but remain consistent with a spinel-like framework; no second phase was assigned in the refinement. (6) Minimal decisive test: Li1.45Mn0.45Sc0.55Cl4 at 500 $^\circ$C keeps the system off 1:1 (to frustrate ordering observed at 1:1) while increasing Li-vacancy content relative to Li1.6. If the abnormal high $\sigma$ arises from the synergy of (i) suppressed 1:1 ordering at 500 $^\circ$C and (ii) elevated vacancy concentration, then moving from Sc = 0.4 → 0.55 at 500 $^\circ$C should push $\sigma$ upward toward the Li1.4 level. Decision rule: • Support: $\sigma$RT $\geq$ 8e-05–1e-04 S/cm with XRD dominated by spinel (small misfits acceptable; no clear emergence of a new ordered phase). • Refute: $\sigma$RT $\leq$ 5e-05 S/cm and/or appearance of multiple new unindexed low-angle peaks suggesting ordering, indicating Li1.4’s high $\sigma$ is not a general off-1:1/vacancy effect at 500 $^\circ$C.
\end{itemize}

\subsection*{\ce{Li_{1.45}Mn_{0.51}Sc_{0.51}Cl_{4}} (Index\# 342)}
\begin{itemize}
\item \textbf{Synthesis temperature}: 500~$^\circ$C
\item \textbf{Ionic conductivity}: $1.20 \times 10^{-4}$~S/cm
\item \textbf{XRD results}: Li3ScCl6 (C12/m1), 86.2 wt\%; Sc7Cl12 (R-3), 13.8 wt\%
\item \textbf{Submitted by}: Abnormal

\item \textbf{Justification}: Step-by-step reasoning from the data and basic principles:
1) Abnormal data pattern: The 1:1 composition Li1.5Mn0.5Sc0.5Cl4 shows $\sigma$RT = 2.2e-06 S/cm at 500 $^\circ$C, a sharp drop vs. its 400 $^\circ$C (4.63e-05) and 450 $^\circ$C (6.43e-05) results. At 500 $^\circ$C, nearby compositions are much higher: Li1.6Mn0.6Sc0.4Cl4 → 3.45e-05 S/cm; Li1.45Mn0.45Sc0.55Cl4 → 1.44e-04 S/cm. Thus, the $\sigma$ collapse is specific to the 1:1 Mn:Sc sample at 500 $^\circ$C.
2) XRD cues at 500 $^\circ$C (1:1): The pattern refines as spinel (Fd-3m) but shows multiple extra\_peaks in the calculated model (low-angle $\sim$15.93–21.77$^\circ$ 2$\theta$). This means the standard disordered Fd-3m model predicts peaks that are not observed, indicating a structural deviation from the disordered reference consistent with altered cation arrangement/intensity distribution at 500 $^\circ$C. In contrast, the 450 $^\circ$C 1:1 sample shows no extra/missing peaks.
3) Basic materials principle: In spinel-type frameworks, higher Li/M sublattice disorder and an optimized population of Li vacancies generally lower migration barriers, while cation ordering (especially near equimolar ratios) can reduce the number of equivalent Li sites and disrupt 3D percolation, raising the activation barrier. This is consistent with the observed $\sigma$ drop only at 1:1 and the recovery when deviating from 1:1 at 500 $^\circ$C.
4) Why introduce slight M-excess at fixed Mn:Sc=1:1: The general formula allows M1±x; increasing M content (x\textgreater0) a little should (by charge neutrality) reduce Li content (increase Li-vacancy concentration) and frustrate any 1:1-optimized Li/M ordering motif. This directly tests whether the 500 $^\circ$C $\sigma$ collapse is tied to an ordering tendency specific to a perfectly balanced 1:1 cation ratio on the octahedral network.
5) Charge neutrality check to set stoichiometry: For Mn2+/Sc3+ mixture at 1:1 with total M=1.02 (Mn0.51Sc0.51), total M charge is 0.51*2 + 0.51*3 = 2.55. To balance 4 Cl--, Li must be 4 -- 2.55 = 1.45. Hence Li1.45Mn0.51Sc0.51Cl4 is the charge-balanced target. Relative to Li1.5Mn0.5Sc0.5Cl4, this increases Li vacancies by $\Delta$y$\approx$0.05 and slightly increases total M occupancy (x$\approx$+0.02), both of which should destabilize long-range cation order.
6) Temperature choice: 500 $^\circ$C is kept intentionally to reproduce the condition under which the $\sigma$ collapse occurs at 1:1; any improvement upon this small off-stoichiometry at the same temperature would isolate the ordering/vacancy mechanism.
Decision rule (within available tools): Success = (i) $\sigma$RT rises well above the collapsed 2.2e-06 S/cm, ideally $\geq$3e-05 S/cm (comparable to off-1:1 neighbors at 500 $^\circ$C), and (ii) the Rietveld fit to Fd-3m shows fewer low-angle extra\_peaks than the 1:1 @500 $^\circ$C sample, indicating a closer match to a disordered spinel. Failure ($\sigma$ remains $\lesssim$1e-05 S/cm with similar Fd-3m mismatches) would argue that the collapse is not primarily driven by 1:1-enabled ordering and would point toward other factors (e.g., different defect populations or microstructure) for follow-up.
\end{itemize}

%% file: auto-gen-si/li1p8mn1p1cl4.tex
\subsection*{\ce{Li_{1.8}Fe_{0.8}In_{0.2}Cl_{4}} (Index\# 20)}
\begin{itemize}
\item \textbf{Synthesis temperature}: 450~$^\circ$C
\item \textbf{Ionic conductivity}: $2.15 \times 10^{-5}$~S/cm
\item \textbf{XRD results}: spinel (Fd-3m), 85.6 wt\%; LiCl (Fm-3m), 14.4 wt\%
\item \textbf{Submitted by}: Initial test

\end{itemize}

\subsection*{\ce{Li_{2}FeCl_{4}} (Index\# 28)}
\begin{itemize}
\item \textbf{Synthesis temperature}: 450~$^\circ$C
\item \textbf{Ionic conductivity}: $1.14 \times 10^{-5}$~S/cm
\item \textbf{XRD results}: spinel (Fd-3m), 100.0 wt\%
\item \textbf{Submitted by}: Initial test

\end{itemize}

\subsection*{\ce{Li_{2}MnCl_{4}} (Index\# 29)}
\begin{itemize}
\item \textbf{Synthesis temperature}: 400~$^\circ$C
\item \textbf{Ionic conductivity}: $5.95 \times 10^{-6}$~S/cm
\item \textbf{XRD results}: spinel (Fd-3m), 87.6 wt\%; LiCl (Fm-3m), 12.4 wt\%
\item \textbf{Submitted by}: Initial test

\end{itemize}

\subsection*{\ce{Li_{1.8}Fe_{1.1}Cl_{4}} (Index\# 194)}
\begin{itemize}
\item \textbf{Synthesis temperature}: 500~$^\circ$C
\item \textbf{Ionic conductivity}: $4.30 \times 10^{-5}$~S/cm
\item \textbf{XRD results}: spinel (Fd-3m), 95.2 wt\%; LiCl (Fm-3m), 4.9 wt\%
\item \textbf{Submitted by}: Abnormal

\item \textbf{Justification}: Objective: discriminate whether high $\sigma$ in Li1.8Fe0.8In0.2Cl4 arises from intrinsic vacancy/disorder (y$\approx$0.2) or undetected Li–In grain-boundary phases. Strategy: replicate the same Li-vacancy concentration (y=0.2) without any trivalent dopant by using Fe2+ excess (x=+0.1) to enforce charge-balanced Li deficiency (y=2x). If vacancy-enabled percolation is intrinsic, Li1.8Fe1.1Cl4 should also show elevated $\sigma$; if the Li–In boundary phase is required, $\sigma$ should fall back to typical Fe–spinel values ($\sim$1e-6–1e-5 S/cm). Temperature 500 $^\circ$C is chosen based on prior Fe–spinel syntheses (450–500 $^\circ$C) to promote phase formation while avoiding halide loss; monitor XRD for FeCl2--its absence supports a clean control.
\end{itemize}

\subsection*{\ce{Li_{1.8}Mn_{1.1}Cl_{4}} (Index\# 250)}
\begin{itemize}
\item \textbf{Synthesis temperature}: 400~$^\circ$C
\item \textbf{Ionic conductivity}: $1.75 \times 10^{-5}$~S/cm
\item \textbf{XRD results}: spinel (Fd-3m), 100.0 wt\%
\item \textbf{Submitted by}: Pattern

\item \textbf{Justification}: Patterns learned: (1) Excess‑metal, vacancy‑rich Fe spinels deliver the highest $\sigma$ in the dataset: Li1.7Fe1.15Cl4 = 6.16e-05 S/cm and Li1.8Fe1.1Cl4 = 4.3e-05 S/cm. (2) Mn is compatible with fast spinel networks when combined with Fe at elevated T (e.g., Li1.6Fe0.8Mn0.4Cl4 at 500 $^\circ$C = 2.97e-05 S/cm; Li1.8Fe0.9Mn0.2Cl4 = 1.44e-05 S/cm). How this design follows those patterns: Li1.8Mn1.1Cl4 transfers the metal‑excess/high‑vacancy strategy to a Mn‑only host to test whether the high‑$\sigma$ window (y $\approx$ 0.2) is host‑agnostic. Charge neutrality: 1.8(Li+) + 1.1$\cdot$Mn2+ = 1.8 + 2.2 = 4+. The aim is to realize a disordered inverse spinel on the cubic Cl sublattice with a percolating Li network similar to the Fe‑only high‑$\sigma$ cases, but without aliovalent dopants that can drive ordering. Why uncertainty remains: The dataset contains no Mn‑only Li2--yMn1+xCl4 examples; phase stability and cation distribution for a Mn‑only inverse spinel are unverified here, and MnCl2 or ordered derivatives could form and suppress $\sigma$. Hence the extrapolation from Fe to Mn carries risk.
\end{itemize}